\documentclass[twocolumn,preprint2]{aastex63}

\usepackage{breakurl}
\usepackage{amsmath,amssymb,xspace}
\usepackage{fp}
\usepackage{mathtools}
\DeclarePairedDelimiter{\abs}{\lvert}{\rvert}

 \newcommand{\gal}{NGC\,4258\xspace}

\defcitealias{Beaton_2016}{Paper~I}
\defcitealias{Hatt_2017}{Paper~II}
\defcitealias{Jang_2018}{Paper~III}
\defcitealias{Hatt_2018a}{Paper~IV}
\defcitealias{Hatt_2018b}{Paper~V}
\defcitealias{Hoyt_2019}{Paper~VI}
\defcitealias{Freedman_2019}{Paper~VIII}
\defcitealias{prop13691}{CCHP}

 \newcommand{\hst}{\emph{HST}\xspace}

\newcommand{\sigmasmooth}{$\sigma_s$\xspace}
\newcommand{\ho}{$H_{0}$\xspace}

\newcommand{\maserdistmod}{29.397} 
\newcommand{\maserdistmodstaterr}{0.024} 
\newcommand{\maserdistmodsyserr}{0.022}
\newcommand{\maserdistmodwerr}{$\mu_0 = \maserdistmod \pm \maserdistmodstaterr $ (stat) $\pm \maserdistmodsyserr$ (sys)\xspace}

\newcommand{\Iextinction}{0.025}
\FPeval{\IextinctioROUNDED}{round(\Iextinction,3)}
\FPeval{\Iextinctionerr}{\Iextinction/2}
\FPeval{\IextinctionerrROUNDED}{round(\Iextinction/2,3)}
\FPeval{\IextinctionerrTOTAL}{round((\Iextinctionerr^2+0.01^2)^0.5,3)}

\newcommand{\trgblum}{-4.054}
\newcommand{\trgblumstaterr}{0.022} 
\newcommand{\trgblumsyserr}{0.039}

\FPeval{\apinffoursevenfive}{round(0.100013,3)} 
\FPeval{\apinffivefivefive}{round(0.0964472,3)} 
\FPeval{\apinfsixzerosix}{round(0.0952613,3)}   
\FPeval{\apinfeightonefour}{round(0.0976345,3)} 
\newcommand{\ZPerr}{0.02}
\newcommand{\EEerr}{0.02}
\newcommand{\Apcorrerr}{0.01}
\newcommand{\sigmasmoothval}{0.11}
\newcommand{\trgbobsval}{25.372}
\newcommand{\trgbobsvalstaterr} {0.014} 
\newcommand{\trgbobsvalsyserr} {0.005}
\newcommand{\trgbcolselerr}{0.01}
\newcommand{\trgbsmoothselerr}{0.01}
\newcommand{\photchoiceerr}{0.02}

 \FPeval\trgbobsvalROUNDED{round(\trgbobsval,3)} 
 \FPeval\trgbredcorrval{round(\trgbobsval-\Iextinction,3)}
 \FPeval\truetrgbdmod{round(\trgbobsval-\Iextinction-\trgblum,2)}
 \FPeval\truetrgbdmodMpc{round(10^(\truetrgbdmod/5)/100000,2)}
 \FPeval\trgbobsvalstaterrROUNDED{round(\trgbobsvalstaterr,3)} 
 \FPeval\trgbobsvalsyserrROUNDED{round((\trgbobsvalsyserr^2+\ZPerr^2+\EEerr^2+\Apcorrerr^2+\trgbcolselerr^2+\trgbsmoothselerr^2+\photchoiceerr^2)^0.5,3)} 
\FPeval\trgbcorrsvalsyserrROUNDED{round( (\trgbobsvalsyserr^2+\ZPerr^2+\EEerr^2+\Apcorrerr^2+\trgbcolselerr^2+\trgbsmoothselerr^2+\photchoiceerr^2 +(\IextinctionerrTOTAL)^2 )^0.5,3)}
 \FPeval\trgbzero{round( \trgbobsval-\Iextinction-\maserdistmod, 3) }
 \FPeval\trgbzerostaterr{round((\trgbobsvalstaterr^2 + \maserdistmodstaterr^2)^0.5, 3)}
 \FPeval\trgbzerosyserr{round((\trgbobsvalsyserrROUNDED^2 + \IextinctionerrTOTAL^2 + \maserdistmodsyserr^2)^0.5, 3)}
 \FPeval\trgbzerototalerr{round((\trgbzerostaterr^2 + \trgbzerosyserr^2)^0.5, 3)}
 
 \FPeval\dmodcombinedstaterr{round((\trgbobsvalstaterr^2+\trgblumstaterr^2)^0.5,3) }
 \FPeval\dmodcombinedsyserr{round((\trgbobsvalsyserr^2+\trgblumsyserr^2+\ZPerr^2+\EEerr^2+\Apcorrerr^2+\trgbcolselerr^2+\trgbsmoothselerr^2+\photchoiceerr^2+\Iextinctionerr^2)^0.5, 3) }

 \FPeval\truetrgbdmodMpcupperrdiststat{ 10^( (\truetrgbdmod+\dmodcombinedstaterr) /5)/100000 }
 \FPeval\truetrgbdmodMpclowerdiststat{ 10^( (\truetrgbdmod-\dmodcombinedstaterr) /5)/100000 }
 \FPeval\truetrgbdmodMpcstaterr{ round( 0.5*(\truetrgbdmodMpcupperrdiststat - \truetrgbdmodMpclowerdiststat) ,2) }
 \FPeval\truetrgbdmodMpcupperrdistsys{ 10^( (\truetrgbdmod+\dmodcombinedsyserr) /5)/100000 }
 \FPeval\truetrgbdmodMpclowerdistsys{ 10^( (\truetrgbdmod-\dmodcombinedsyserr) /5)/100000 }
 \FPeval\truetrgbdmodMpcsyserr{ round( 0.5*(\truetrgbdmodMpcupperrdistsys - \truetrgbdmodMpclowerdistsys) ,2) }

 \newcommand{\trgbredcorrvalwerr}{$\mathrm{F814W}_{0}=\trgbredcorrval\pm \trgbobsvalstaterrROUNDED$~(stat)~$\pm \trgbcorrsvalsyserrROUNDED $~(sys)$~\mathrm{mag}$\xspace}

 \newcommand{\trgbzeropoint}{$M^{\rm TRGB}_{\rm F814W} = \trgbzero \pm \trgbzerostaterr$~(stat) $\pm \trgbzerosyserr$~(sys)\xspace}
 
\graphicspath{{./}{figures/}}

\shorttitle{CCHP IX. The TRGB in NGC\,4258}
\shortauthors{CCHP team}

\begin{document}

\title{The Carnegie-Chicago Hubble Program. IX. \\ Calibration of the Tip of the Red Giant Branch Method \\ in the Mega-Maser Host Galaxy, \gal\ (M106) \footnote{Based on observations made with the NASA/ESA Hubble Space Telescope, obtained at the Space Telescope Science Institute, which is operated by the  Association of Universities for Research in Astronomy, Inc., under NASA contract NAS 5-26555. These observations are associated with programs GO-9477 and GO-10399}}


\correspondingauthor{In Sung Jang}
\email{hanlbomi@gmail.com}

\author{\bf In Sung Jang}
\affil{Leibniz-Institut f\"{u}r Astrophysik Potsdam (AIP), An der Sternwarte 16, 14482 Potsdam, Germany}
\affil{Dept. of Astronomy \& Astrophysics, Univ. Chicago, 5640 S. Ellis Ave., Chicago, IL, ~~60637} 

\author{\bf Taylor Hoyt}
\affil{Dept. of Astronomy \& Astrophysics, Univ. Chicago, 5640 S. Ellis Ave., Chicago, IL, ~~60637} 

\author{\bf Rachael Beaton}
\affil{Dept. of Astrophysical Sciences, Princeton University, 4 Ivy Lane, Princeton, NJ ~~08544}
\affil{The Observatories, Carnegie Institution for Science, 813 Santa Barbara St., Pasadena, CA ~~91101}

\author{\bf Wendy L. Freedman}
\affil{Dept. of Astronomy \& Astrophysics, Univ. Chicago, 5640 S. Ellis Ave., Chicago, IL, ~~60637}

\author{\bf Barry F. Madore} 
\affil{The Observatories, Carnegie
Institution for Science, 813 Santa Barbara St., Pasadena, CA ~~91101}
\affil{Dept. of Astronomy \& Astrophysics, Univ. Chicago, 5640 S. Ellis Ave., Chicago, IL, ~~60637}

\author{\bf Myung Gyoon Lee}
\affil{Department of Physics \& Astronomy, Seoul National University, Gwanak-gu, Seoul 151-742, Republic of Korea}
\author{\bf Jillian R. Neeley}
\affil{Department of Physics, Florida Atlantic University, 777 Glades Rd., Boca Raton, FL 33431, USA}
\author{\bf Andrew J. Monson}
\affil{Department of Astronomy \& Astrophysics, The Pennsylvania State University, 525 Davey Lab, University Park, PA 16802, USA}
\author{\bf Jeffrey A. Rich }
\affil{The Observatories, Carnegie
Institution for Science, 813 Santa Barbara St., Pasadena, CA ~~91101}

\author[0000-0002-1143-5515]{\bf Mark Seibert} 
\affil{Unaffiliated, Pasadena, CA 91101, USA}

\begin{abstract}
In the nearby galaxy \gal, the well-modeled orbital motion of $\rm H_2O$ masers about its supermassive black hole provides the means to measure a precise geometric distance. As a result, NGC 4258 is one of a few ``geometric anchors’’ available to calibrate the true luminosities of stellar distance indicators such as the Tip of the Red Giant Branch (TRGB) or the Cepheid Leavitt law.
In this paper, we present a detailed study of the apparent magnitude of the TRGB within \gal\ using publicly-available \hst\ observations optimally situated in the unreddened stellar halo along the minor axis, spanning distances ranging from 8 to 22~kpc in projected galactocentric radius, or $6\arcmin$ (13~kpc) to $30\arcmin$ (66~kpc) in distance along the semi-major axis.
 We undertake a systematic evaluation of the uncertainties associated with measuring the TRGB in this galaxy,  based on an analysis of  54 arcmin$^2$ of HST/ACS imaging.
 After quantifying these uncertainties, we measure the TRGB in \gal to be \trgbredcorrvalwerr.
Combined with a recent 1.5\% megamaser distance to NGC 4258, we determine the absolute luminosity of the TRGB to be \trgbzeropoint mag.
This new calibration agrees to better than 1\% with an independent calibration presented in Freedman et al. (2019, 2020) that was based on detached eclipsing binaries (DEBs) located in the LMC.
\end{abstract}
\keywords{distance scale --- stars: Population II --- galaxies: individual (NGC4258) --- galaxies: stellar content --- galaxies: structure}
\section{Introduction}\label{sec:intro}

There have been major advances in our ability to measure cosmological parameters to high precision and accuracy over the past two decades \citep[e.g.,][and references therein for an early review]{FreedmanTurner2003}. Effectively unchanged, however, is the role of the Hubble Constant (\ho) in framing our current cosmological understanding. Early universe and local distance-scale methods appear to disagree at a significant level,  causing  concern that our cosmological model may require revision \citep[see, e.g.,][and references therein]{Freedman_2017, Verde2019}. 
An equally valid interpretation of our current \ho-tension is that the quoted uncertainties have been underestimated, not due to a lack of rigor in analyses, {\it per se}, but rather because not all sources of systematic uncertainty have yet been identified, as the quoted uncertainties have  decreased from the 10\% to the 1\% level over the last two decades. 

The nearby spiral galaxy, \gal\ plays a significant role in establishing the absolute calibration of the modern distance scale. At a distance of 7.58~Mpc \citep{Reid_2019}, \gal is the closest galaxy, beyond the Local Group, with a geometric distance measurement; and, it is still sufficiently nearby that detailed studies of its resolved stellar populations can be made by currently-operating, space-based telescopes.
The next closest megamaser host galaxy is UGC~3789,  at a distance of $D = 51.5^{+4.5}_{-4.0}$ Mpc or $\mu$ = 33.6$^{+0.19}_{-0.17}$ mag \citep[][]{Pesce_2020}. 
The data analysis techniques employed in determining the megamaser-based distances are both elegant and complex  \citep[for example, see {\it The Introduction to the Megamaster Cosmology Project},][]{Reid2009}. 
Once the maser distance is secured, other (more widely applicable) distance indicators that have also been studied in NGC~4258, such as the TRGB stars and Cepheids, can have their zero points accurately established.

In addition to playing an important role in the distance ladder, \gal\ has been included in several large-scale galaxy evolution programs that have provided a rich set of multi-wavelength datasets \citep[][]{Heald_2011,Kim_2011, Spencer_2014,Merritt_2016,Sabbi_2018}.
Thus, there is a wealth of archival data on this galaxy, much of it yet to be exploited for the purposes of calibrating the extragalactic distance scale. Here we tap into that reservoir of data, specifically in the context of calibrating the TRGB method.

In the following, we use \hst\ imaging of \gal\ as a means of identifying and quantifying the uncertainties associated with TRGB measurements. In the process, we provide a high-confidence TRGB measurement in \gal\, and rigorously quantify the associated systematic and statistical uncertainties.

This paper is organized as follows:
First in \autoref{sec:data}, we describe the \hst\ imaging datasets that we have analyzed, describe the image processing undertaken, followed by an account of stellar photometry methods used and their calibration. In \autoref{sec:validation}, we present a detailed comparison of two independent photometry analyses (using DOLPHOT and DAOPHOT), allowing us to examine the impact of differences in adopted input parameters. In \autoref{sec:trgb}, we present the CCHP TRGB measurement, followed in \autoref{sec:trgbcompare} by a discussion of previously published measurements. Calibration of the TRGB using the maser distance is presented in \autoref{sec:geocalib}. The results of the paper are summarized in \autoref{sec:summary}.
Finally, in a set of appendices, 
we discuss 
the photometric accuracy with the choice of sky fitting algorithms (\autoref{app:appa}), 
stellar-population variations and their potential impact (\autoref{app:appb}), and 
provide an independent analysis of earlier WFPC2 data (\autoref{app:appc}).

\section{Data and Data Processing}\label{sec:data}

\subsection{Archival HST Data}\label{sec:archival_data}

\begin{figure*}
    \centering
    \includegraphics[width=1.0\textwidth]{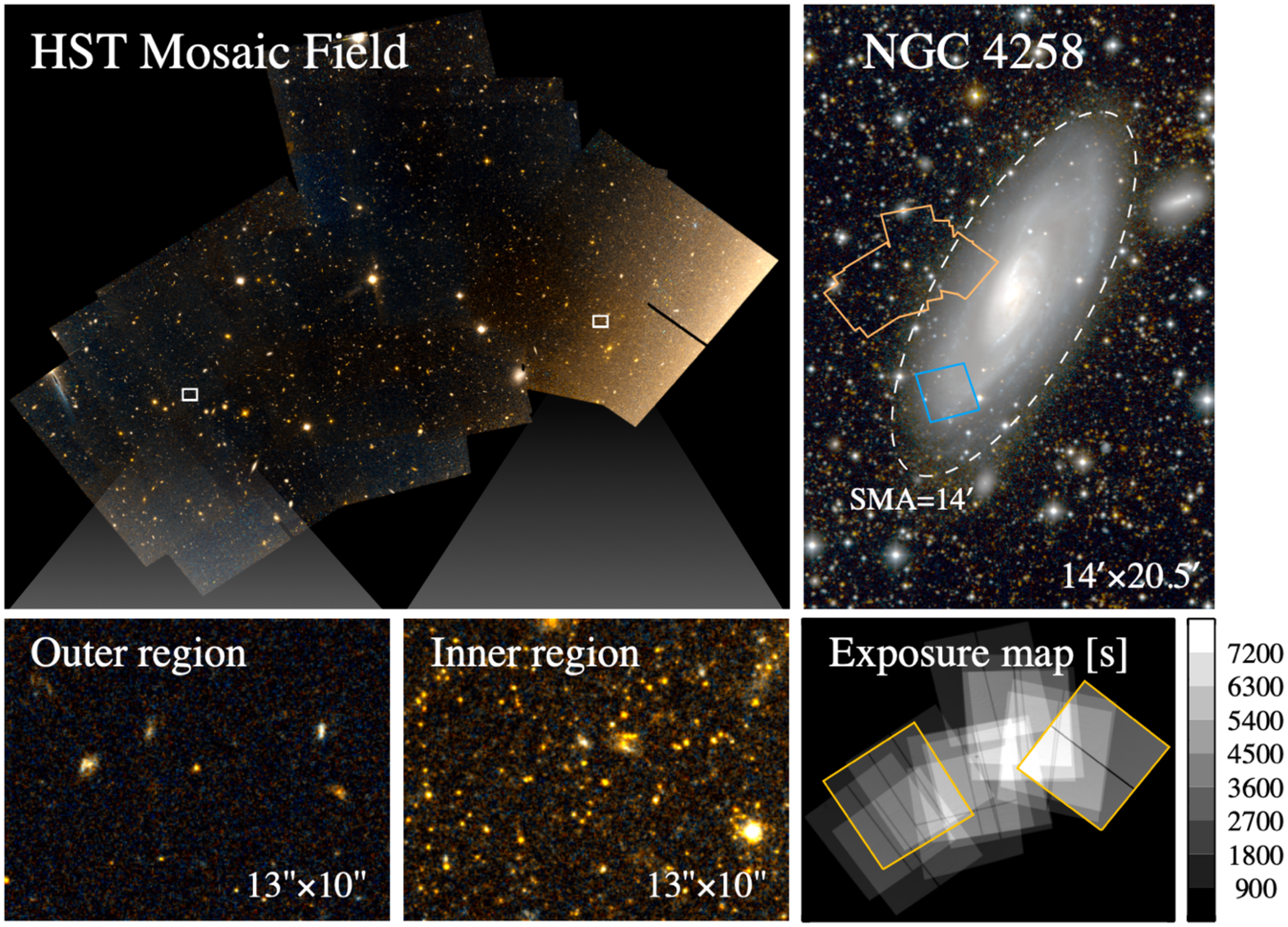}
    \caption{
    Summary of the \hst\ imaging data for NGC\,4258 used in this study. 
    Top left: The full mosaic of 15 individual \hst/ACS fields. 
    The area of the mosaic is 54 arcmin$^2$, approximately five times larger than a single $ACS$ field, and spans from $SMA = 6\arcmin$ (13~kpc) to $30\arcmin$ (66~kpc) from the center of \gal. 
    Top right: identification of the $HST/ACS$ fields used in this and previous studies overlaid on an image taken with the Dragonfly array \citep[][]{Abraham_2014}.
    The mosaic $ACS$ field (orange polygon) lies on the minor axis of the galaxy and a single field (blue square) in the disk of the galaxy indicates observations taken for the analysis of Cepheid variables, which we later use for comparison with the halo mosaic dataset.
    We use the region beyond SMA = 14$\arcmin$ to define a selection of halo member stars.
    Bottom: zoom-in views of the outer (left) and inner (middle) regions of the mosaic $ACS$ field. The exposure weight map in $F814W$ is shown in the last panel.
    Yellow squares indicate Field~1 (right) and Field~13 (left) we used for a photometry comparison.
    North is up and east is to the left in all the panels.
     }
    \label{fig:imagedata}
\end{figure*}

\autoref{fig:imagedata} presents the totality of mosaiced data used for this work as retrieved from the MAST archive, largely constructed from ACS observations obtained in parallel-mode. 
The upper left panel of \autoref{fig:imagedata} is a color image of the contiguous area used for this work that has both F555W and F814W imaging to a sufficient depth for a TRGB analysis; red, green, and blue channels of the color image are taken from F814W, (F555W--F814W)/2, and F555W, respectively. 
The mosaic is made from 15 individual ACS pointings with details on the observations for each pointing provided in \autoref{tab:data_table}.
As given in \autoref{tab:data_table}, the individual pointings span a region ranging from a galactocentric radius of $4\farcm2$ to $10\farcm5$ in projection from the center of \gal.
\gal\ has an inclination of $i$ = 68.3$^{\circ}$ with a position angle of 150$^{\circ}$ 
\citep[taken from HyperLeda\footnote{\url{http://leda.univ-lyon1.fr/}};][]{Makarov_2014}, such that our mosaic corresponds to a disk semi-major axis (SMA) ranging from $5\farcm7$ (13~kpc) to $31\farcm1$ (68~kpc).

The upper right panel of \autoref{fig:imagedata} places this imaging in the context of the large-scale structure of \gal, using a wide-field image taken with the {\it Dragonfly} array \citep[][]{Abraham_2014,Merritt_2016}, which is constructed from the $g$ and $r$--band image data with $r$ for red, $g$ for green, and $2\times g-r$ for blue.
The region we use is outlined in yellow and is more-or-less on the minor axis of \gal; the majority of its area is beyond a de-projected semi-major axis (SMA) of 14$'$, corresponding to a physical radius in the \gal-disk of approximately 30 kpc. 
The blue outlined region (to the lower left) interior to SMA = 14$'$ is the ACS pointing that has been used in prior work to measure the TRGB \citep[e.g,][]{Macri_2006, Rizzi_2007, Reid_2019} that we will refer to as the Disk pointing.

Recent work in the Milky Way has used chemo-dynamical evidence to identify disk-like stars as far as $R_{\rm Gal} = $24~kpc, while also finding stars with disk-like chemistry at significant vertical scale heights \citep[][and references therein]{Hayes_2018}. 
Thus, we anticipate that our innermost regions may be contaminated by stellar populations belonging to the outer disk of \gal, while our outermost regions are expected to have significantly less disk contamination.

The lower left panels of \autoref{fig:imagedata} give a sense of how the source density varies across the region, by providing zooms into two locations of the mosaic that are representative of lower stellar density (outer region) and higher stellar density (inner region). 
As is visually apparent in these panels, the region spans a large range of on-sky source densities. 
The inclusion of high-density regions in our image footprint is intentional; these regions allow us to study explicitly how source density, and variations in the underlying stellar populations, impact measurements of the TRGB.

Lastly, the lower right panel of \autoref{fig:imagedata} provides the exposure map of the image mosaic for the F814W filter.
In the mosaic, the median exposure time is 2,780~sec with a standard deviation of 1,770~sec;
the maximum exposure time is 10,460~sec.
Using prior work as a guide, the magnitude of the TRGB is expected to be around F814W = 25.3-25.4~mag \citep[e.g.,][]{JangLee_2017}; the median exposure time predicts a signal-to-noise of 17.1 to 18.6 at this magnitude, enabling the detection and measurement of the TRGB across the full extent of the mosaic.

\begin{deluxetable*}{lccccrrcccr} 
\tabletypesize{\small}
\setlength{\tabcolsep}{0.05in}
\tablecaption{Archival $HST/ACS$ Images used in This Study \label{tab:data_table} }
\tablewidth{0pt}

\tablehead{\colhead{Field} & \colhead{R.A.} & \colhead{Decl.} & \colhead{$R$} & \colhead{$SMA^a$} & \multicolumn{2}{c}{Exposure Time(s)} & \colhead{Obs. date} & \multicolumn{2}{c}{Zero-points$^{c}$} & \colhead{Prop. ID}  \\
& (2000.0) & (2000.0) & [arcmin]& [arcmin (kpc)$^b$] & $F555W$  & $F814W$ & & $F555W$ & $F814W$ & }

\startdata
F1 & 12:19:18.31 & 47:20:09.8   & 4.2 & 11.0 (24.3) & 5700  & 2600  & 2003-05-05    & 25.736 & 25.531 & 9477 \\
F2 & 12:19:25.32 & 47:20:32.3   & 5.3 & 14.1 (31.1) & 900   & 900   & 2005-07-30    & 25.731 & 25.528 & 10399 \\ 
F3 & 12:19:26.11 & 47:20:11.5   & 5.3 & 14.0 (30.8) & 900   & 900   & 2005-07-24    & 25.731 & 25.528 & 10399 \\ 
F4 & 12:19:31.26 & 47:21:10.9   & 6.5 & 17.3 (38.2) & 900   & 900   & 2005-07-08    & 25.731 & 	25.528 & 10399 \\ 
F5 & 12:19:32.63 & 47:21:06.7   & 6.7 & 17.8 (39.2) & 900   & 900   & 2005-07-11    & 25.731 & 25.528 & 10399 \\ 
F6 & 12:19:33.13 & 47:21:36.3   & 7.0 & 18.7 (41.1) & 900   & 900   & 2005-07-15    & 25.731 & 25.528 & 10399 \\ 
F7 & 12:19:38.43 & 47:22:22.0   & 8.2 & 21.8 (48.1) & 900   & 900   & 2005-07-01    & 25.731 & 25.528 & 10399 \\
F8 & 12:19:38.50 & 47:19:30.5   & 7.1 & 18.1 (39.9) & 900   & 900   & 2005-07-02    & 25.731 & 25.528 & 10399 \\
F9 & 12:19:38.73 & 47:19:17.7   & 7.1 & 18.0 (39.6) & 900    & 900   & 2005-06-29    & 25.731 & 25.528 & 10399 \\
F10 & 12:19:41.91 & 47:19:25.9   & 7.7 & 19.4 (42.8) & 900   & 900   & 2005-07-02    & 25.731 & 25.528 & 10399 \\
F11 & 12:19:42.48 & 47:18:30.1   & 7.7 & 18.7 (41.3) & 900   & 900   & 2005-04-02    & 25.732 & 25.528 & 10399 \\
F12 & 12:19:51.59 & 47:19:30.6   & 9.3 & 23.5 (51.8) & 900   & 900   & 2005-06-01    & 25.731 & 25.528 & 10399 \\
F13 & 12:19:55.78 & 47:18:56.7   & 10.0 & 24.5 (54.0) & 900   & 900   & 2005-06-02    & 25.731 & 25.528 & 10399 \\
F14 & 12:19:53.91 & 47:17:33.7   & 9.6 & 22.2 (48.9) & 900   & 900   & 2005-05-28    & 25.732 & 25.528 & 10399 \\
F15 & 12:19:59.20 & 47:18:00.5   & 10.5 & 24.8 (54.7) & 900   & 900   & 2005-05-25    & 25.732 & 25.528 & 10399 \\
\enddata
\tablenotetext{a}{Semi-Major Axis of the stellar disk that has a position angle of 150$^{\circ}$ and an inclination angle of 68.3$^{\circ}$ (HyperLeda).}
\tablenotetext{b}{Assuming the distance of $D$=7.58 Mpc from \citet{Reid_2019}.}
\tablenotetext{c}{From the ACS zero-point calculator (\url{https://acszeropoints.stsci.edu/}).}
\end{deluxetable*} 


\subsection{Photometry}\label{sec:photometry}

The first step in our analysis is the construction of a photometric catalog of the resolved stars in the halo of NGC~4258. We now briefly describe that process.

In this study, we have undertaken the bulk of the photometric reductions using DOLPHOT \citep{Dolphin_2000}, as it has distortion correction routines for the HST instruments that are not available within our existing DAOPHOT pipeline.
The archival imaging data for \gal have large offsets with different orientation angles between the fields and were not taken with a standard dither pattern, as for our CCHP program. One of the main difficulties encountered in the \gal\ data reduction involved the  alignment and registration of the individual FLC images, each of which has strong geometric distortions known to be generated by the camera optics.

The CCHP has developed its own point-spread-function (PSF) fitting photometry pipeline that is based on DAOPHOT \citep{stetson_1987} and models synthetic PSFs from TinyTim \citep{Krist_2011}. 
The most detailed description of the pipeline is given in \citet{Beaton_2019} and the pipeline is specifically designed to provide robust, homogenous measurements of resolved stars in HST images.
It has been used to measure TRGB distances to nearby SN~Ia host galaxies \citep{Hoyt_2019,Freedman_2019,Beaton_2019}.
The CCHP pipeline was designed around single fields and not multiple fields with differing orientations, as can be seen in  \autoref{fig:imagedata}.
While the pipeline could be adapted, we found the native tools of DOLPHOT better adapted to the task.
With these caveats between the two reduction packages noted, we performed extensive tests to search for and quantify differences between the methodologies used in this paper and the CCHP pipeline results.

We developed an independent photometry pipeline based on DOLPHOT to perform simultaneous photometry on all the individual image frames associated with the mosaic data of \gal. 
We have also reduced the data for a subset of the fields using DAOPHOT.
This approach provides a unique opportunity to check the robustness of our photometry specifically with respect to the choice of software, as discussed in detail in \autoref{sec:dol_dao}.

DOLPHOT uses the WCS information contained in image headers to obtain an initial alignment solution. 
We inspected the header WCS entries to see whether the images were properly aligned, and found that some image frames showed visible offsets of up to $2\arcsec$ with respect to others.
Such WCS-offsets are naturally expected for images taken from different visits that use different guide stars for alignment.
We thus aligned images by updating the header WCS information using the  DrizzlePac.TweakReg task \citep{Gonzaga_2012} as follows.

The TweakReg task in DrizzlePac provides an automated interface for detecting sources and using them to compute the shifts between the input images.
We initially applied this task to the original WCS in the FLC images, but found that the resulting alignment solution was not sufficient to get robust photometry. 
We suspect that this uncertainty is most likely driven by the lack of bright stellar sources in the images for two reasons:  \gal\ is located at a high galactic latitude ($b\sim68.8$ deg) and the target field is sampling the low-surface brightness (halo) regions of the galaxy.
In this situation, the DrizzlePac manual suggests using a list of reliable sources manually selected from initial photometry from DAOPHOT or SExtractor; thus, we carried out preliminary PSF photometry on the FLC images using DAOPHOT to construct just such a list of bright sources.

At this stage, the source list contained a large number of cosmic rays or other defects, in addition to genuine astronomical objects (stars \& background galaxies).
To reduce the non-astronomical contamination, we matched pairs of source lists for co-spatial regions in the WCS domain with a matching tolerance of 1 pixel ($0\farcs05$).
With the source lists cleaned, TweakReg task found a better alignment solution with a typical {\it rms} value of 0.1 pixels ($0\farcs005$).
The well-aligned FLC images were then used to make a stacked drizzled frame with the AstroDrizzle task. 
DrizzlePac also updates the data quality (DQ) extension of FLC images by providing pixel-level flags identifying cosmic rays and hot pixels.
Finally, we repeated the steps from our preliminary PSF photometry, this time using cosmic-ray-masked FLC images to again improve the alignment solution.

A standard sequencing of the DOLPHOT procedures for the ACS module was then executed: 
$acsmask$ -$ splitgroups$ - $calcsky$ - $dolphot$. 
A total 120 FLC images were simultaneously photometered with the TinyTim PSFs implemented within DOLPHOT.
We used the drizzled mosaic frame in the F814W-band to coordinate the reference source positions.
For $calcsky$,  we used the values recommended in the manual: $r_{in}$ = 15, $r_{out}$ = 35, $step$ = 4, $\sigma_{low}$ = 2.25, and $\sigma_{high}$ = 2.00.
A list of input parameters we used for the DOLPHOT task is summarized in \autoref{tab:params0}. 
These parameters are broadly consistent with the recommended values in the manual, but there are slight changes.
For example, we set {\tt Align} = 4 with {\tt UseWCS} = 1 to achieve a more elaborate geometric distortion correction using a full third-order polynomial fit.
The {\tt FitSky} parameter determines the algorithm used for the local background estimation. 
Following the photometry processing in the PHAT survey \citep{Dalcanton_2012}, we adopted {\tt FitSky} = 3, which fits simultaneously the sky background and the PSF within the fitting radius.
Most importantly, we set {\tt ApCor} = 0, which means that we did not apply the aperture corrections automatically determined from the code itself; instead we examined bright stars and determined the correction values manually. 
The details of our manually-derived aperture correction are given in the next section.

We later explore the robustness of our photometry by varying several key parameters ({\tt RPSF}, {\tt FitSky}, and {\tt ACSpsfType}) in \autoref{sec:validation}.
From these analyses, we find that the systematic error associated with the choice of photometry parameters is at most 0.01~mag.

\begin{figure}[h]
    \centering
   \includegraphics[width=0.45\textwidth]{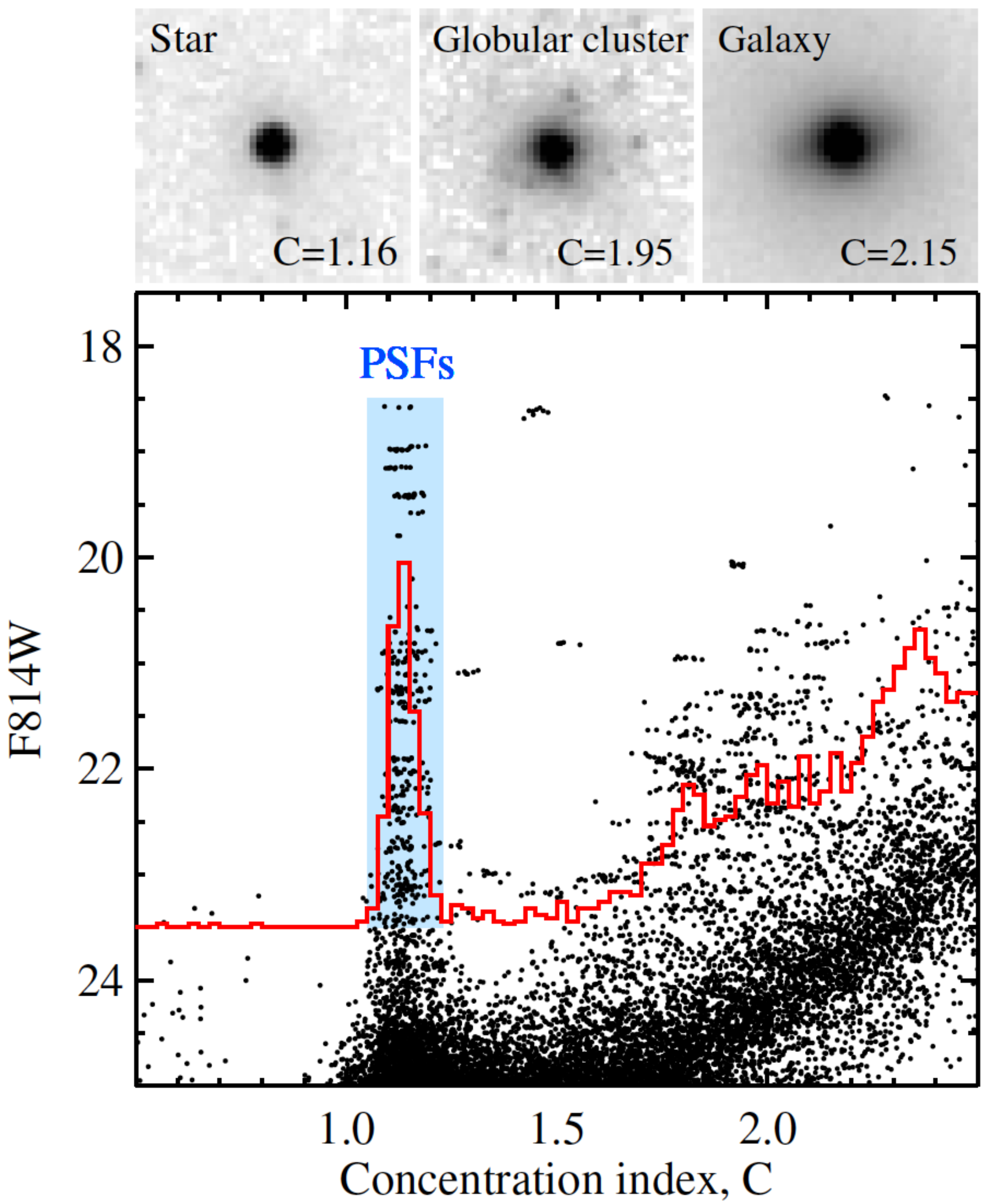}
    \caption{
        Selection of high-quality point sources for the aperture correction. 
        The {\it Concentration Index}, $C$, the difference in magnitude measured with small and large aperture radii, more specifically: $C$ = mag($r=0.8$ pix) -- mag($r=2.5$ pix).
        A strong plume of point sources is seen at $C\sim1.1$. 
        Overlaid in red is a histogram for the sources brighter than F814W = 23.5~mag, showing the steep rise of sources at larger $C$.
        The upper three panels (from left to right) show the F814W-band thumbnail images ($2\arcsec \times2\arcsec$) for a stellar point source, a globular cluster, and a background galaxy, with their $C$ values in F814W marked at the bottom of each image.
        These selection criteria were only applied for the determination of aperture corrections.
    }
    \label{fig:apcor}
\end{figure}

\begin{deluxetable*}{lll} 
\tabletypesize{\small}
\setlength{\tabcolsep}{0.05in}
\tablecaption{DOLPHOT Processing Parameters \label{tab:params0}}
\tablewidth{0pt}
\tablehead{\colhead{Description} & \colhead{Parameter} & \colhead{Value}   \\ } 

\startdata
Inner and outer sky radii   & img\_apsky    & 15 25 \\
Photometry apeture size     & img\_RAper    & 8     \\
$\chi$--statistic aperture size & img\_RChi & 2     \\
Photometry type             & PSFPhot       & 1     \\   
Fit sky?                    & FitSky        & 3     \\  
PSF size                    & img\_RPSF      & 13    \\
Spacing for sky measurement & SkipSky       & 2     \\
Sigma clipping for sky      &  SkySig       & 2.25  \\
Second pass finding stars   & SecondPass    & 5     \\   
Searching algorithm         & SearchMode    & 1     \\   
Sigma detection threshold   & SigFind       & 2.5   \\   
Multiple for quick-and-dirty photometry & SigFindMult & 0.85 \\    
Sigma output threshold      & SigFinal      & 3.5   \\  
Maximum iterations          & MaxIT         & 25    \\    
Noise multiple in imgadd    & NoiseMult     & 0.10  \\   
Fraction of saturate limit  & FSat          & 0.999 \\        
Find/make aperture corrections? & ApCor     & 0     \\  
Force type 1/2?             & Force1        & 0     \\  
Use WCS for initial alignment? & useWCS     & 1     \\  
Align images?               & Align         & 4     \\  
Allow cross terms in alignment? & Rotate    & 1     \\    
Centroid box size           & RCentroid     & 2     \\  
Search step for position iterations & PosStep& 0.25 \\  
Maximum single step in position iterations  & dPosMax & 3.0 \\  
Minimum separation for two stars for cleaning & RCombine & 1.415    \\   
Minimum S/N for PSF parameter fits  & SigPSF & 3.0  \\   
Make PSF residual image?    & PSFres        & 1     \\        
Coordinate offset           & psfoff        & 0.0   \\     
Use saturated cores?        & FlagMask      & 4     \\      
Use the DOLPHOT CTE correction  & ACSuseCTE & 0     \\ 
PSF Type                    & ACSpsfType    & 0     \\
\enddata
\end{deluxetable*} 

\subsection{Aperture Correction and Photometry Calibration}\label{sec:apcor}

The approximate nature of PSF photometry, together with the finite size of the radius used by crowded-field photometry applications, requires an additional term for calibration, commonly known as the `aperture correction'.
This term corrects for the difference between the measured flux at finite radius and the total stellar flux, as measured at infinity.
As previously described, we performed PSF photometry on our dataset, and so the infinite aperture correction step is divided into two distinct steps: a correction from PSF to aperture magnitudes at finite radius, and a correction from the finite radius aperture to infinity. The latter is provided for ACS by STScI, so the bulk of this section covers the former, namely our empirical determination of the transformation from the PSF to an aperture magnitude system at finite radius.

DOLPHOT provides a routine for the aperture correction by setting {\tt ApCor} = 1.
This routine selects bright stars in the individual frame images and calculates the necessary correction.
This option can be convenient in generic cases, however it can be difficult to know exactly what goes into the automated process, and for the high precision, high accuracy photometry demanded by the extragalactic distance scale, we prefer to directly measure the aperture corrections outside of DOLPHOT.  
Accordingly, we independently measured the aperture correction (with  {\tt ApCor} = 0) by first manually examining reliable point sources. 

We selected the bright point sources based on the concentration parameter, $C$, which is the difference between magnitudes measured with small and large aperture radii: $C$ = mag($r$ = 0.8 pix) -- mag($r$ = 2.5 pix).
\autoref{fig:apcor} shows our selection criteria for these bright point sources.
The $C$ values are derived from the aperture photometry on individual FLC images, after the cosmic ray masking. 
We plotted the values taken from all the images together in \autoref{fig:apcor}.
Because we have 120 FLC images in the mosaic field, some stars were multiply imaged onto different positions on the detector and independently photometered.
We expect intrinsic scatter in the $C$ values, as the PSF is known to vary across the HST focal plane.
Nevertheless, the figure shows a strong plume of sources at $C\sim1.1$, which are identified as \textit{bona fide} point sources.
In addition to the empirical determination of the point source sequence given above,
we applied this same {\it Concentration Index} procedure to artificial point sources that were injected into several regions across the FLC images. 
We confirmed that the injected sources, and thus our PSF model, also has a mean $C$ value of $\sim 1.1$, consistent with that seen in the real data.

In the upper panels in \autoref{fig:apcor}, we display thumbnail images of three representative types of sources taken from the stacked F814W frame: a star, a globular cluster candidate in the \gal\ halo\footnote{The position of the globular cluster candidate is 
 RA = 12:19:25.89, and DEC = +47:22:55.5. It is $6\farcm72$ away from the center of \gal\, corresponding to a projected physical distance of 14.8~kpc at 7.58~Mpc \citep{Reid_2019}.}, and a background galaxy. 
It is clear that the star ($C\sim1.16$) is well separated from the other detected sources ($C\sim1.95$ for the globular cluster candidate and $C\sim2.15$ for the background galaxy), in both the {\it Concentration Index} and their morphology.
We thus selected the bright point sources in the shaded region of the figure, and used them for the aperture correction after a careful visual inspection.

Using this list of bright and isolated point sources (standard stars), we computed growth curves for the standard stars contained in each FLC frame. 
We set 12 aperture radii to construct the growth curves out to 14 pixels ($0\farcs7$).
For a small subset of the selected standard stars, the growth curves provided evidence that additional faint neighbor stars were contaminating the aperture. 
To correct for this contamination, and to measure the unbiased transformation from PSF to aperture magnitudes, we manually subtracted the neighbor sources using the SUBSTAR task in DAOPHOT, and then re-computed the growth curves from the cleaned stellar profile.
In 
\autoref{tab:apcor}, we provide the final list of stars used for the aperture corrections and the calibrated magnitudes we measure at a 5 pixel radius.
Photometric zero-points are derived from the online STScI ACS Zeropoints Calculator\footnote{https://acszeropoints.stsci.edu/} as listed in \autoref{tab:data_table}.
There are 64 and 38 standard stars distributed over the entire mosaic of the F814W and F555W fields, respectively.
The mean photometric error of the standard stars is smaller than 0.01~mag in each filter.

From the growth curves, a correction from the PSF magnitudes to the $r=5$ pixel aperture magnitudes was determined for each frame. 
We found that each F814W and F555W frame contains on average $\sim$7 and $\sim$4 standard stars, respectively.
The statistical uncertainty for the aperture corrections is smaller than 0.01 mag for F814W and at the level of 0.01 mag for F555W. 
From this, we adopt 0.01~mag as a systematic uncertainty associated with the aperture correction.
The mean correction for the 60 F814W (F555W) frames is 0.077 (0.089)~mag with a standard deviation of 0.012 (0.019)~mag.
The calibrated magnitudes in each FLC frame were then combined in the flux domain, resulting in the deepest possible photometric catalog.
With the PSF photometry now calibrated to a 5-pixel aperture, a correction from r = 5 pixel to infinity \citep{Bohlin_2016} was applied: --0.1726 mag for F814W and --0.1537 mag for F555W (all data were obtained before the ACS servicing mission).

\begin{deluxetable*}{lcccccc} 
\tabletypesize{\footnotesize}
\setlength{\tabcolsep}{0.05in}
\tablecaption{A List of Bright and Isolated Stars used for the Aperture Correction  \label{tab:apcor} }
\tablewidth{0pt}

\tablehead{\colhead{ID} & \colhead{R.A.} & \colhead{Decl.} & \colhead{F814W} & \colhead{N(stdev)$^{a}$}  & \colhead{F555W} & \colhead{N(stdev)$^{a}$} \\
& (2000.0) & (2000.0) &  ($r=5$ pix) &   &    ($r=5$ pix) & }
\startdata
  1        &  12:19:08.518     &  47:19:52.95     &  $22.396 \pm  0.007$   &    2(0.014) &  ...   &  ... \\
  2        &  12:19:08.558     &  47:20:09.50     &  $19.797 \pm  0.001$   &    2(0.001) &  ...   &  ... \\
  3        &  12:19:08.580     &  47:19:46.43     &  $23.096 \pm  0.016$   &    1        &  $23.753 \pm  0.013$   &      1  \\
  4        &  12:19:09.346     &  47:20:05.46     &  $22.926 \pm  0.009$   &    2(0.008) &  $22.997 \pm  0.007$   &      1  \\
  5        &  12:19:11.656     &  47:21:03.28     &  $21.106 \pm  0.004$   &    1        &  ...   &  ... \\
  6        &  12:19:11.741     &  47:20:25.83     &  $22.682 \pm  0.011$   &    1        &  ...   &  ... \\
  7        &  12:19:11.825     &  47:19:17.62     &  ...                   &       ...   &  $23.796 \pm  0.013$   &      1  \\
  8        &  12:19:12.235     &  47:18:51.94     &  $22.706 \pm  0.013$   &    1        &  ...   &  ... \\
  9        &  12:19:15.970     &  47:21:16.93     &  $22.504 \pm  0.005$   &    5(0.019) &  ...   &  ... \\
 10        &  12:19:16.128     &  47:19:02.56     &  $20.468 \pm  0.002$   &    2(0.000) &  $21.762 \pm  0.004$   &      1  \\
 11        &  12:19:17.761     &  47:19:23.40     &  $23.100 \pm  0.014$   &    1        &  ...   &  ... \\
 12        &  12:19:18.153     &  47:20:53.39     &  $23.223 \pm  0.015$   &    1        &  ...   &  ... \\
 13        &  12:19:18.318     &  47:22:00.09     &  $23.411 \pm  0.011$   &    2(0.021) &  ...   &  ... \\
 14        &  12:19:18.933     &  47:18:12.73     &  $22.960 \pm  0.009$   &    2(0.004) &  ...   &  ... \\
 15        &  12:19:20.303     &  47:18:19.42     &  ...                   &         ... &  $20.490 \pm  0.005$   &      1  \\
 16        &  12:19:20.969     &  47:20:46.56     &  $22.027 \pm  0.007$   &    4(0.024) &  $23.924 \pm  0.012$   &      1  \\
 17        &  12:19:21.504     &  47:19:27.83     &  $21.904 \pm  0.005$   &    4(0.003) &  ...   &  ... \\
 18        &  12:19:22.507     &  47:21:36.83     &  $22.758 \pm  0.007$   &    5(0.015) &  ...   &  ... \\
 19        &  12:19:22.991     &  47:20:20.11     &  ...                   &         ... &  $23.530 \pm  0.016$   &      3(0.036) \\
 20        &  12:19:22.994     &  47:20:20.05     &  $22.367 \pm  0.005$   &    8(0.010) &  ...   &  ... \\
\enddata
\tablenotetext{a}{The number of independent measurements from different FLC frames, and their standard deviation of the $r = 5$~pixel aperture  magnitudes.}
\tablenotetext{}{(This table is available in its entirety in machine-readable form.)}
\end{deluxetable*} 


\begin{figure*}
\centering
    \includegraphics[width=0.8\textwidth]{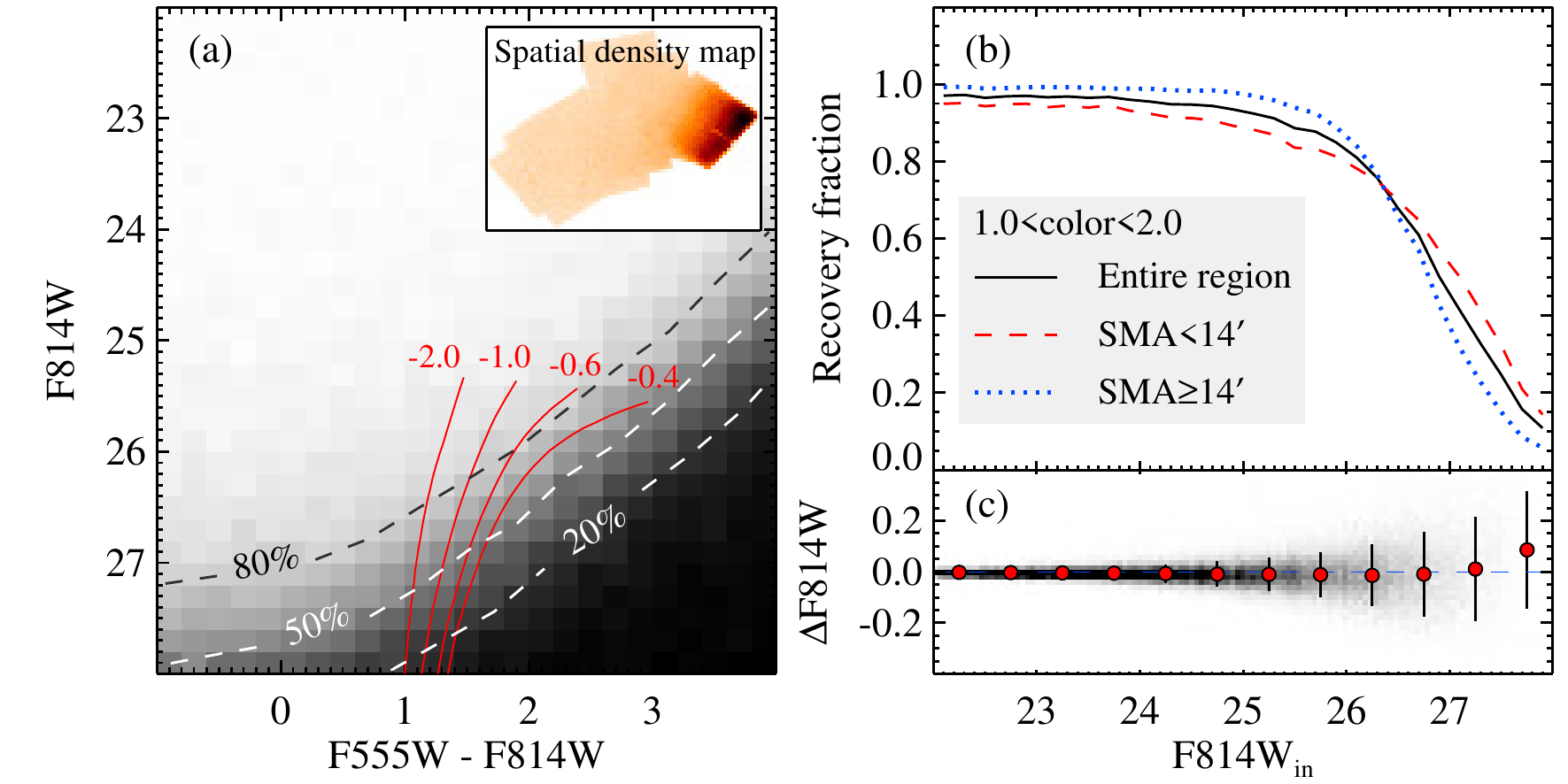}
    \caption{
    Photometric completeness determined from artificial star tests.
    (a) Color Magnitude Diagram (CMD) displaying photometric completeness using a color coding from black for fully incomplete (0\%) to white (100\%) for fully complete. 
    Three representative completeness levels (20\%, 50\%, and 80\%) are shown by dashed lines.
    Red lines indicate the 10 Gyr isochrones with [Fe/H] = --2.0, --1.0, --0.6 and --0.4 dex in the Padova model at the distance of \gal\ \citep{Bressan_2012}.
    An inset panel displays the spatial density map of the injected artificial stars, which were placed to mimic the true spatial distribution of real stars. 
    (b) Recovery fractions as a function of input F814W magnitude (F814W$_{in}$) for the stars with $1.0 < \rm{F555W - F814W} < 2.0$ for the full mosaic and regions interior and exterior to SMA=14$\arcmin$ (solid, dashed, and dotted lines, respectively).
    The 50\% recovery limit occurs at $\sim$27~mag, approximately 1.6 mag fainter than the expected magnitude of the TRGB.
    (c) The difference between the input and measured magnitudes, $\Delta F814W = F814W_{in} - F814W_{out}$, as a function of F814W$_{\rm in}$ for the full mosaic and stars with $1.0 < \rm{F555W - F814W} < 2.0$. }
    \label{fig:art_stars}
\end{figure*}

\subsection{Artificial Star Tests}\label{sec:art_tests}
A robust determination of stellar flux and associated properties requires in-depth understanding of uncertainties.
In crowded-field photometry, it is known that the error of stellar flux is not simply defined by purely Poisson statistics because unresolved sources below the detection limit will contribute to the 
main source flux, potentially resulting in a bias.
There are three key indicators to assess the robustness of stellar photometry: statistical errors (precision), systematic errors (accuracy), and recovery rates (completeness).
Estimating them using the real star photometry alone is very difficult (or near to impossible); rather a series of tests with artificial stars is needed.
We thus carried out extensive artificial star tests on our photometry of \gal.

We used DOLPHOT for the artificial star tests with the general procedures described in the DOLPHOT User's guide and, in particular, the manual provided by Bill Harris\footnote{http://physwww.mcmaster.ca/$\sim$harris/dolphot\_primer.txt}.
We generated artificial stars with a wide range in color ($-1<(\rm{F555W}-\rm{F814W})<4$~mag) and in magnitude (22 $<\rm{F814W}<28$ mag) to span the color-magnitude range shown in \autoref{fig:art_stars}(a).
The spatial distribution of the input stars was based on the distribution of real stars, placing more stars in the inner high-surface brightness regions (\autoref{fig:art_stars}(a) inset).
About 350,000 artificial stars were injected into each image and recovered alongside the real stars using the identical procedures as adopted to derive the primary catalog.
Because DOLPHOT performs the test in a star-by-star manner, the intrinsic degree of stellar crowding local to the artificial star is not significantly impacted by this process.

The results of the artificial star tests are shown in \autoref{fig:art_stars}.
\autoref{fig:art_stars}(a) displays the recovery rate in color-magnitude space compared to stellar isochrones from PARSEC \citep{Bressan_2012}. 
\autoref{fig:art_stars}(b)  displays the recovery fraction of the artificial stars with $1.0 < \rm{F555W-F814W} < 2.0$, 
as a function of input F814W magnitude (F814W$_{in}$) for the full mosaic (solid black) and the regions interior to (red dashed) and exterior to (blue dotted) SMA=$14\arcmin$.
The recovered stars have passed the point source selection criteria that we shall explain in the next section.
We found that our photometry is complete enough to detect the \gal\ TRGB: the recovery fractions are higher than 90\% at the anticipated TRGB at F814W $\sim$ 25.4~mag.
The fractions are still as high as 80\% at approximately 1 mag fainter than the TRGB.
We also found that there is a difference in completeness depending on the spatial selection.
For faint sources (with $\rm{F814W} \gtrsim 26.5$ mag), the inner crowded region (SMA $< 14\arcmin$) shows higher recovery fractions than the outer region (SMA $> 14\arcmin$).
We infer that this is likely due to the difference in the effective exposure time, as the inner SMA $< 14\arcmin$ region has a longer median exposure time (F814W = 3017s and F555W = 6165s) than the outer region (F814W = 2736s and F555W = 2742s).
By contrast, for brighter sources ($24 \lesssim \rm{F814W} \lesssim 26$ mag), the trend reverses, suggesting that crowding effects, due to higher source densities present in the inner region, could be suppressing the observed recovery fraction (see \autoref{fig:imagedata}).

Representative differences between the input and output photometry of the entire mosaic field are shown in \autoref{fig:art_stars}(c).
The median and standard deviation of the difference in input and output magnitudes for artificial stars is given for 0.5 mag bins. 
The standard deviation, which is an indicator of the true statistical error, ranges from $\sigma_{F814W}\sim0.01$~mag to $\sim$0.14~mag.
At the anticipated TRGB magnitude, the photometric dispersion is estimated to be $\sigma_{F814W} = \pm0.08$ mag, indicating that the quality of our photometry is sufficient to make a high-precision TRGB measurement.
Turning our interest to the systematic errors, we also confirm that our photometry is accurate.
The median offset between the input and output photometry is only $\Delta F814W \sim0.01$ mag at the expected TRGB magnitude of $F814W \simeq 25.4$~mag.
This is much smaller than the statistical errors at the same magnitude, and also smaller than the typical TRGB detection error in our previous studies ($\sigma_{F814W} \sim \pm0.04$ mag).
Here we emphasize that our TRGB detection technique is based on a series of tests using the artificial stars, so any errors in photometry will be properly incorporated in the final error budget.

\begin{figure}
    \centering
    \includegraphics[width=0.42\textwidth]{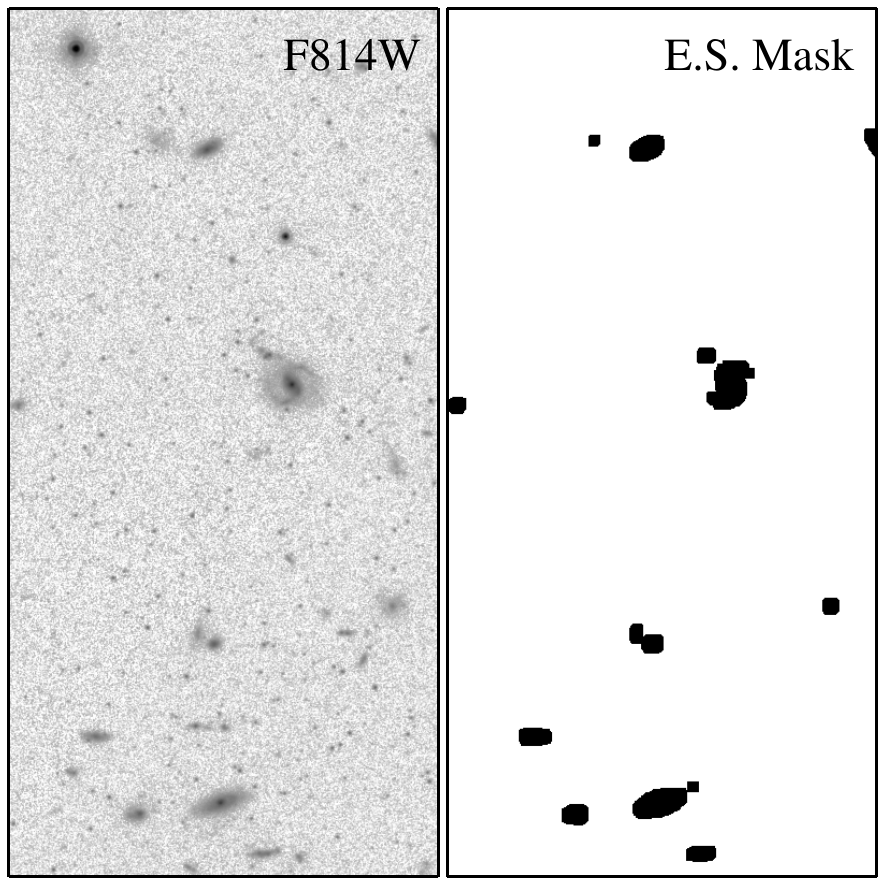}
    \caption{Identification of extended sources.
    SExtractor was run on the stacked F814W image and a mask of extended sources (either $F814W < 24.5$~mag and class $< 0.2$ or $F814W < 18.5$~mag and class $> 0.8$) was created using the segmentation map. 
    Small sections ($10\arcsec \times 20\arcsec$) of the stacked drizzled image (left) and the resulting mask map (right) are shown.
    The DOLPHOT derived sources within the masked black areas were not used any further in the analysis.
    }
    \label{fig:mask}
\end{figure}

\begin{figure*}
    \centering
    \includegraphics[width=0.7\textwidth]{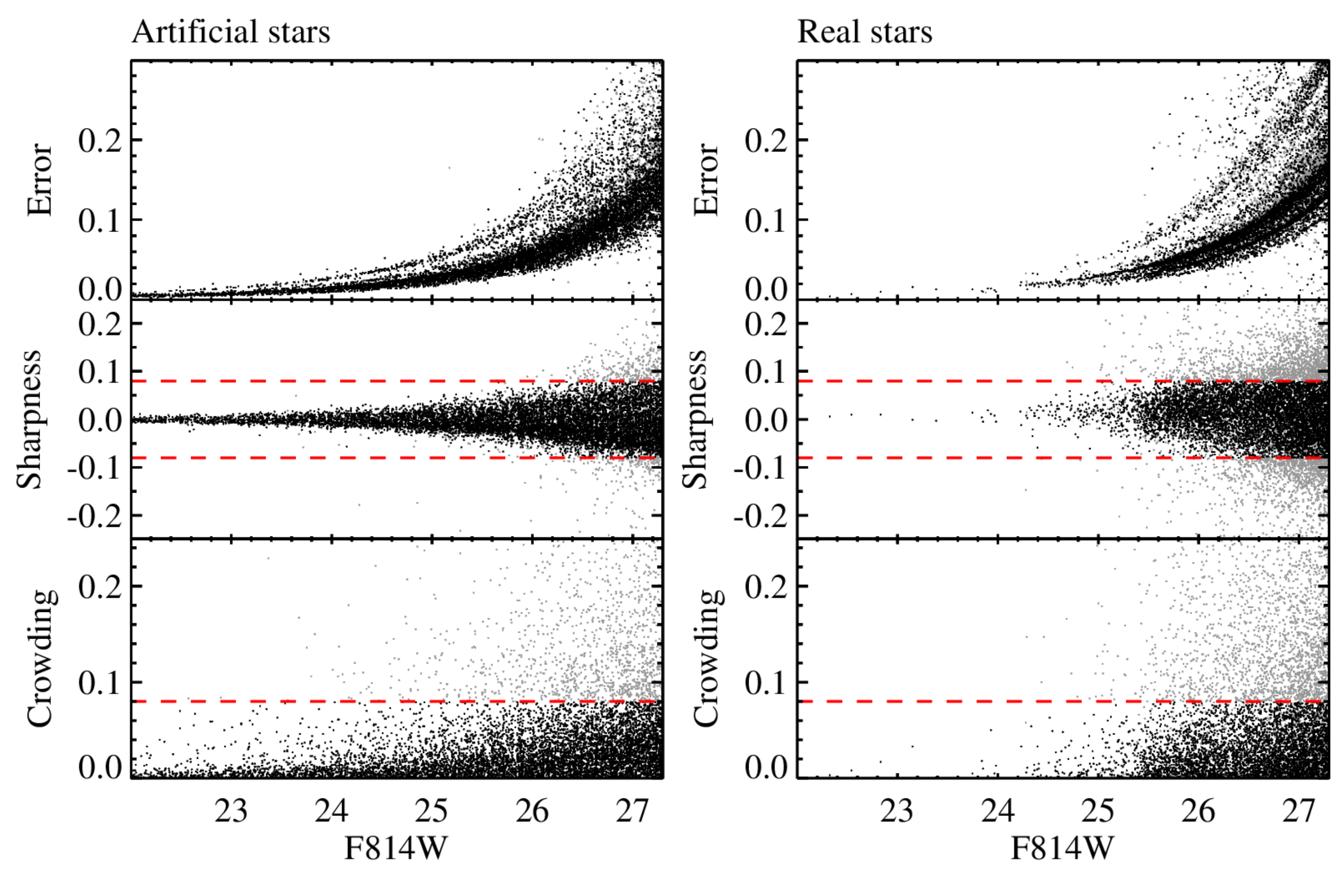}
    \caption{
    Determination of the point source selection criteria.
    Distribution of the 
    error ($\sigma_{\rm F814W}$)
    sharpness (Sharp$_{\rm F814W}$ + Sharp$_{\rm F555W}$) and 
    crowding (Crowding$_{\rm F814W}$ + Crowding$_{\rm F555W}$) 
    parameters as a function of F814W magnitudes are shown.
    The multiple sequences shown in the error distribution plots (top panels) are the results of the different effective exposure times across the full mosaic.
    The point-source selection criteria (dashed lines) were based upon the distributions of artificial stars (left) and applied to the observed sources (right).
    Sources that passed both the sharpness and crowding-based criteria are indicated by black dots. }
    \label{fig:culls}
\end{figure*}

\subsection{Point Source Selection}\label{sec:culls}

The raw DOLPHOT output catalog contains the photometry for various types of objects. 
Selecting reliable point sources from the raw catalog is important because the inclusion of other types of sources will dilute the desired signal coming from the resolved stars.
For the aperture correction, we used the {\it Concentration Index}, $C$, to select bright point sources, but this process also illustrated how $C$ varied with magnitude, largely showing the parameter to be unreliable across the full catalog. 
Thus, we select point sources for our TRGB analysis using a more sophisticated series of selection criteria that can be divided into two main steps:
 1) using an extended source mask, and 
 2) using photometric diagnostic parameters (sharpness and crowding).
The first step is very efficient in rejecting contaminants in the bright magnitude range, where the sequence of point sources is not clearly defined, while the second step is relatively more important at the faint end.
This approach is similar to the method used in the GHOSTS survey \citep{deJong_2007,Radburn-Smith_2011}, which undertook an extensive analysis of the stellar populations in nearby galaxy halos.

To construct the extended source mask, we follow the procedures described in \citet{Radburn-Smith_2011}.
Briefly, we ran SExtractor \citep{Bertin_1996} on the stacked, drizzled F814W image, and obtained a segmentation map with a source catalog.
The resulting segmentation map was used to make region masks, identifing intrinsically extended sources (F814W $<$ 24.5~mag and class $<$ 0.2), as well as for very bright sources, showing extended halos around their PSF core (F814W $<$ 18.5~mag and class $>$ 0.8).
The final mask map was constructed after convolving the region mask with a Gaussian filter having a smoothing scale of 3 pixels. The resulting map is shown in \autoref{fig:mask}.
Masked regions occupied only 1.5\% of the total area of the mosaic field, but, as intended,  they reject a significant number of bright extended sources.
DOLPHOT-derived sources, falling on the white area of the mask map, were carried forward for the further analysis.

The second step for the point-source selection was made using photometric diagnostic parameters.
It is known that the ``sharpness" and ``crowding'' parameters  returned from DOLPHOT are useful in selecting point sources \citep{Dalcanton_2009}.
Our point-source selection criteria are shown in \autoref{fig:culls}. The left three panels show the reported errors, sharpness and crowding values for artificial stars that have the colors of the blue RGB stars, approximately $1.0 < (\rm{F555W-F814W}) < 2.0$~mag.
We used the distributions of these idealized stars as a means of selecting real stars in the observed dataset (seen in the right three panels).
Conservative selection criteria were chosen, as shown by dashed lines in the figure.
Together with the ``sharpness" and ``crowding'' parameters, we also used Type = 1 (clean stars) sources, and required signal-to-noise ratios higher than 3 in both filters.
We provide a summary of the final point-source selection criteria below:

\begin{equation}
    \abs{\mathrm{SHARPNESS_{F555W}} + \mathrm{SHARPNESS_{F814W}}} < 0.08,
\end{equation}
\begin{equation}
    \mathrm{CROWDING_{F555W}} + \mathrm{CROWDING_{F814W}} < 0.08,
\end{equation}
\begin{equation}
    \mathrm{S/N_{F555W} > 3.0}, \qquad \mathrm{S/N_{F814W} > 3.0},  \quad \mathrm{and} \quad \mathrm{Type = 1.}  
\end{equation}

\begin{figure*}[h]
    \centering
    \includegraphics[width=0.890\textwidth]{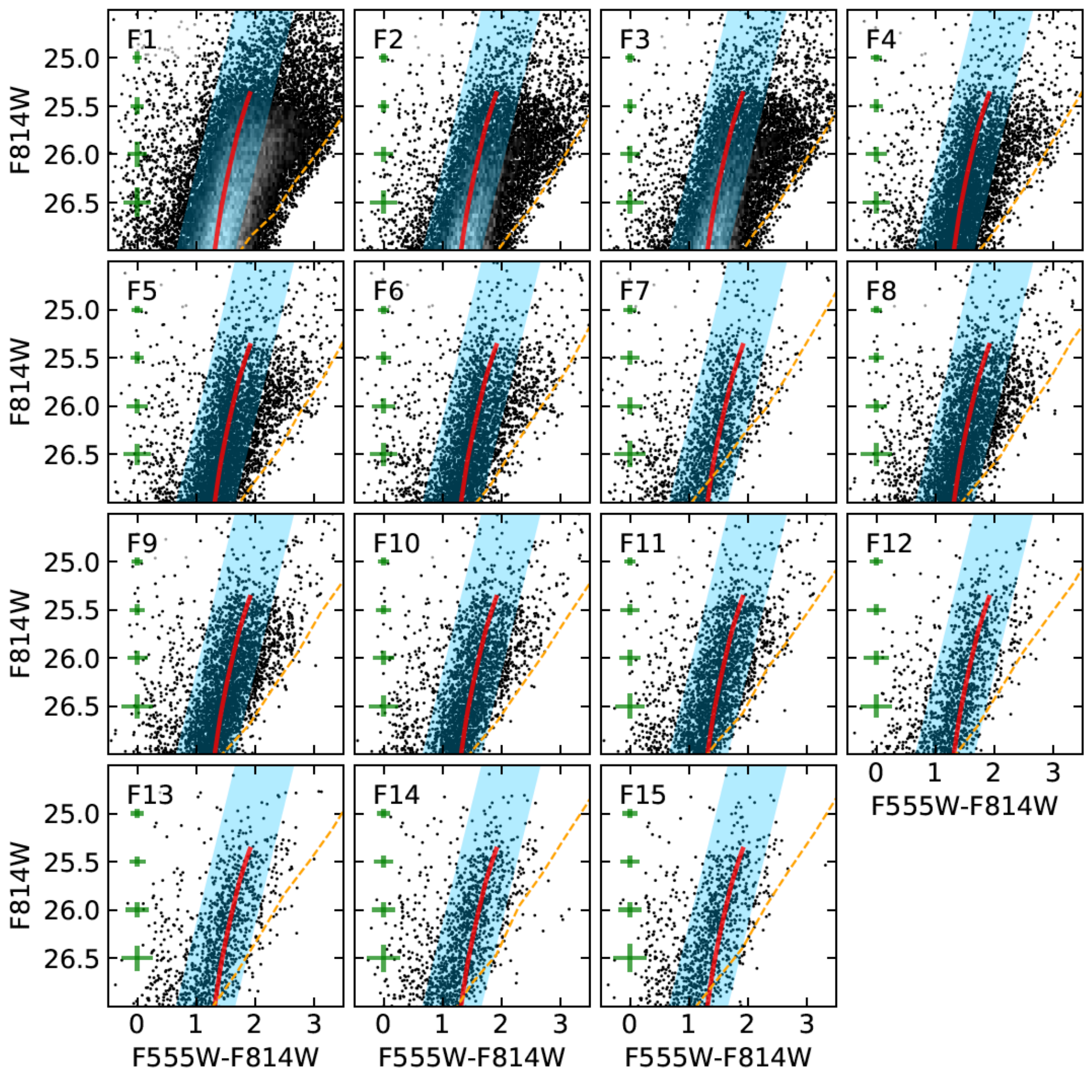}
    \caption{ 
    CMDs in F814W versus F555W -- F814W for each of the 15 ACS pointings in the mosaic field. 
    The blue shaded region is the same in each panel and represents the ``blue TRGB'', which corresponds to the color range where the TRGB absolute magnitude varies little with color. Photometric errors measured at F555W -- F814W $\sim$ 1.5 are marked at F555W -- F814W = 0.
   To guide the eye, in each panel a PARSEC stellar isochrone, with an age of  10 Gyr and [Fe/H] $=-1.0$~dex, is shown by a solid red line. We shifted the isochrone to the distance modulus ($(m-M)_0=\maserdistmod$ \citep{Reid_2019}) and foreground reddening ($A_{F606W}=0.040$ and $A_{F814W}=0.025$ \citep{Schlafly_2011}) of NGC 4258. The 50\% recovery rate is indicated by a yellow dashed line to the right of the RGB.
    }
    \label{fig:cmds_15fields}
\end{figure*}

\autoref{fig:cmds_15fields} presents color-magnitude diagrams (CMDs) of the selected point sources in the 15 ACS pointings.
The panels are ordered approximately by their projected distance from the center of \gal\ (see \autoref{tab:data_table}). 
The shaded cyan-box in each panel of \autoref{fig:cmds_15fields} highlights the ``blue TRGB'' \citep[e.g., that adopted by][among others]{JangLee_2017,Freedman_2019} and the shaded-region is the same for all panels. 
From a visual inspection it is clear that the inner fields (i.e., F1 to F6) contain significant numbers of RGB stars spread to either side of the ``blue TRGB'', whereas the RGB sequences for the outermost fields (i.e., F12 to F15) are well defined and entirely contained within the ``blue TRGB'' region.
Photometric errors are reasonably small in all the fields, indicating that the observed large spread is not simply a result of the higher stellar density in the inner regions.
This is consistent with our assessment that the innermost regions of the mosaic suffer from significant contamination from young, metal-rich stars that populate the outer disk, whereas the outer regions are consistent with being uncrowded and consisting of a  primarily old and  metal-poor stellar population. 

\begin{figure*}
    \centering
    \includegraphics[width=0.8\textwidth]{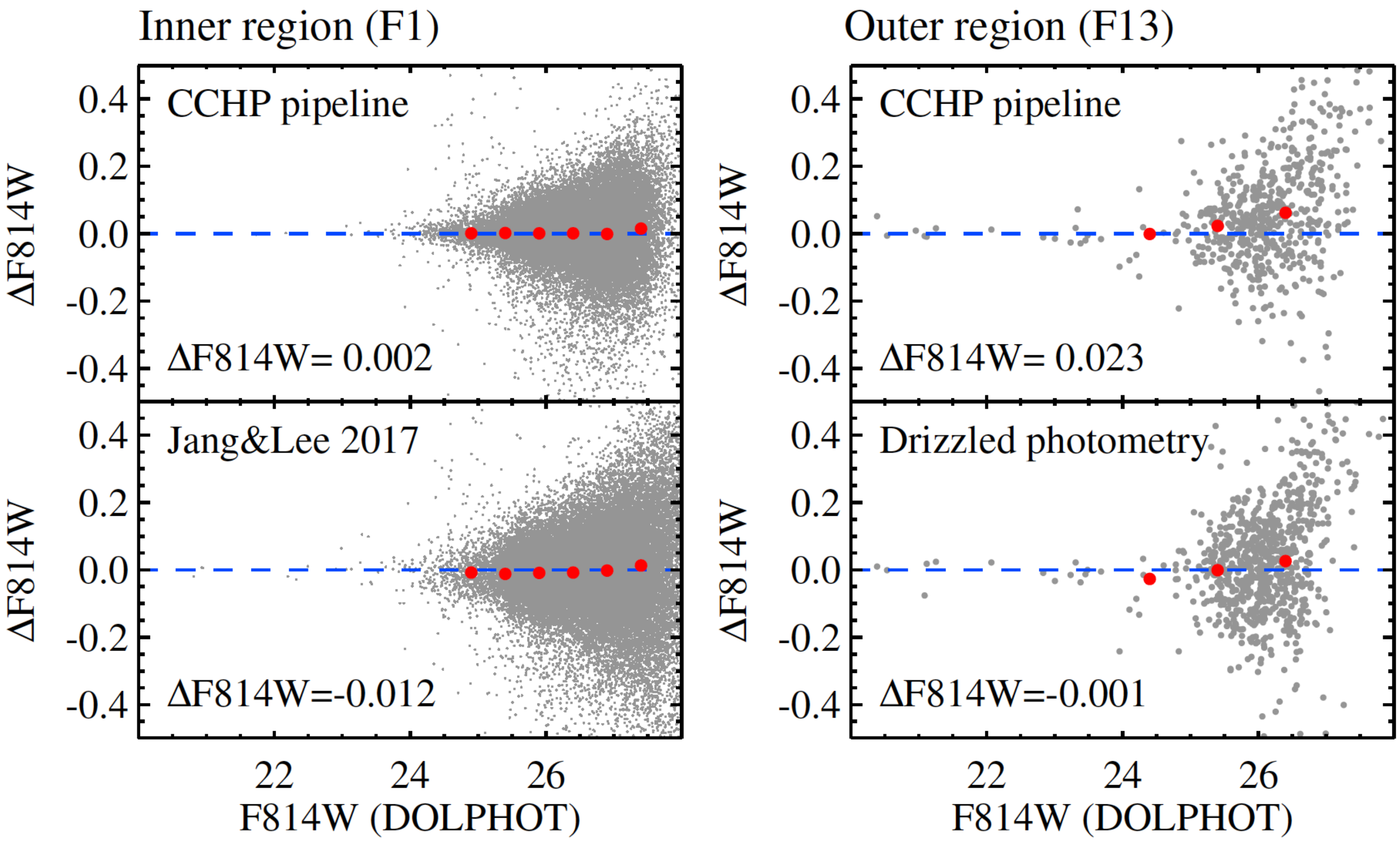}
    \caption{
    Comparison of photometry between the DOLPHOT and the two DAOPHOT-based reductions. 
    We select two fields in our mosaic dataset: F1 in the inner region at $SMA\sim11\arcmin$ (left panels), and F13 in the outer region at $SMA\sim24\farcm5$ (right panels). 
    Comparisons with the CCHP pipeline photometry (DAOPHOT applied to individual FLC frames) are presented in the top.
    Also shown in the bottom are comparisons with the photometry from drizzled images (DAOPHOT applied to stacked frames using Drizzlepac) from \citet{JangLee_2017} (bottom-left) and that produced for this study (bottom-right).
    Median offsets (defined as DOLPHOT minus DAOPHOT) in each magnitude bin are marked by red dots.}
    \label{fig:dol_dao}
\end{figure*}

\section{ Photometry Validation Checks}\label{sec:validation}

In this section, we explore the impact of choices in the photometric processing of our frames on our measurement of the TRGB magnitude in this galaxy.
We first examine methodological concerns by comparing two independent sets of stellar photometry, one  obtained using DOLPHOT-based routines, the other using DAOPHOT (\autoref{sec:dol_dao}).
We then examine details of the DOLPHOT reduction, testing three key parameters: Sky Fitting Methods (in \autoref{sec:fitsky}), PSF Size (in \autoref{sec:psfsize}), and PSF Types (in \autoref{sec:psftype}).

\subsection{DOLPHOT vs. DAOPHOT Comparisons}\label{sec:dol_dao}

In this section, we compare the photometry from two photometric data-reduction packages. As mentioned earlier, in order to process the complex mosaic dataset shown in \autoref{fig:imagedata}, we adopted a DOLPHOT-based methodology unique to this paper, in contrast to the DAOPHOT-based pipeline custom-designed for the CCHP.
In the current section we aim to justify this methodology and ensure that any calibration resulting from it can be safely folded into the \citet{Freedman_2019} sample.

We select two fields, one in the inner region at $SMA\sim11\arcmin$ (F1) of our mosaic and one in the outer region at $SMA\sim24\farcm5$ (F13), as marked by yellow squares in \autoref{fig:imagedata}, within which we compare the DOLPHOT results with the CCHP pipeline photometry \citep[as described in][]{Beaton_2019}. 
In \autoref{fig:dol_dao}, we show a star-by-star comparison of the two independent photometry sets of the inner (top-left) and outer (top-right) regions.
The median offsets at the expected level of the TRGB ($F814W\sim25.4$~mag) are small: $\Delta F814W = 0.002$~mag in the inner region and $\Delta F814W =  0.023$~mag in the outer region. 
The error of the median offsets should be larger than 0.014~mag, considering the aperture correction uncertainty of each reduction (0.01~mag).
The larger offset in the F13 field is likely due to the shorter exposure time (900s, $S/N\sim8$ at $F814W\sim25.4$~mag) than the F1 field (2600s, $S/N\sim16$ at $F814W\sim25.4$~mag), together with the larger uncertainty associated with the small number of stars in the TRGB magnitude bin.
For example, the F13 field has 85 stars that have a standard deviation of 0.089~mag in the $\pm$0.25~mag bin of the expected TRGB level.
In the same magnitude bin, the F1 field has about 30 times more stars ($N = 2702$) with a smaller standard deviation ($0.058$~mag).
Importantly, because the photometry used for our final TRGB measurement  (SMA $\geq 14\arcmin$)
reached a median effective exposure time of 2736s, we expect the bulk of our photometry to exhibit the same degree of excellent agreement observed in the F1 comparison. 
This result shows that our two independent pipelines are both reliable and compatible; a TRGB calibration obtained from our DOLPHOT-based reductions can be confidently applied to the TRGB distances measured in \citet{Freedman_2019}.

In addition to the comparison with the standard CCHP pipeline photometry above, we further investigate the photometry determined from drizzled images.
\citet{JangLee_2017} independently reduced F1 in the inner region of our mosaic field.
They performed point-source photometry using DAOPHOT on the stacked drizzled frames (DRC) using empirical PSFs that were constructed from the same DRC images. 
We compared our DOLPHOT-based photometry with the photometry used in \citet{JangLee_2017} and show the result in the bottom left panel of \autoref{fig:dol_dao}.
The bottom right panel of \autoref{fig:dol_dao} provides the same comparison for F13.
We found that these two independent reductions agree to the 1\% level, the median difference at the TRGB magnitude being $\sim$0.01~mag.

\begin{figure*}
    \centering
    \includegraphics[width=0.8\textwidth]{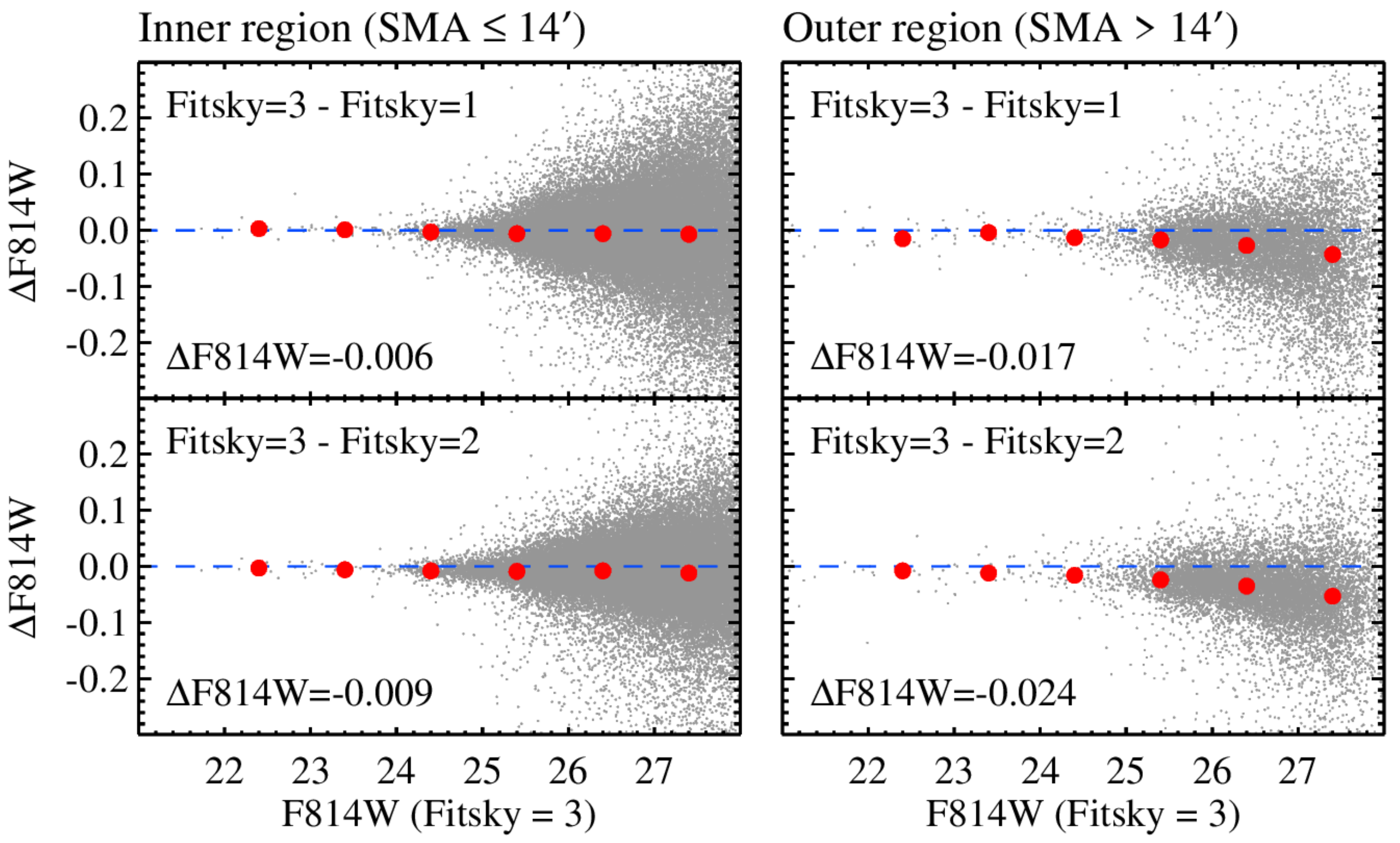}
    \caption{
    Comparison of photometry determined with different sky fitting options in DOLPHOT ({\tt Fitsky} = 1, 2, and 3). 
    The top panels show magnitude differences between the {\tt Fitsky} = 1 and 3 photometry in the inner (left) and outer (right) regions of the mosaic field.
    The bottom panels are the same as top panels except for the {\tt Fitsky} = 2 photometry is used for a comparison.
     Median offsets in each magnitude bin are indicated by red dots.
     The median offset at the anticipated magnitude of the TRGB is also marked in each panel.
    }
    \label{fig:fitsky_real}
\end{figure*}


\subsection{Sky Fitting Methods}\label{sec:fitsky}

An accurate measurement of the local sky background is one of the basic requirements for stellar photometry.
There are several methods (or algorithms) to measure the local sky, but there is not a singular, universally-adopted algorithm that can meet all scientific goals. 
In most cases, we chose one of the possible methods based on our prior experiences and proceeded with the data reduction.
However, there could be systematic effects in the resulting photometry caused by the method chosen for the sky estimation.
We therefore tested our photometry to assess the degree of uncertainty associated with the sky estimation.

There are three representative options in the local sky estimation in DOLPHOT: {\tt Fitsky}  = 1, 2, and 3.
According to the DOLPHOT manual, the {\tt Fitsky}  =1 option measures the local sky background from an annulus (typically 15 and 35 pixels) prior to the PSF fit.
The options {\tt Fitsky}  = 2 and 3 are similar; they include the sky level and the PSF model in a two-component fit to the point source profile.
Both {\tt Fitsky} = 2 and 3 can be used for the crowded field photometry, while {\tt Fitsky} = 1 should only be used for very uncrowded regions.

We reduced the mosaic data with all of these above-mentioned sky fitting options.
Following the manual, we set {\tt RSky} = 4 10 when the {\tt Fitsky}  = 2 option is used.
Similarly, we used a small aperture radius of {\tt RAper} = 3 pixels for the {\tt Fitsky} =1 and 2.
\autoref{fig:fitsky_real} displays a comparison of the three reductions.
Magnitude differences between the {\tt Fitsky}  = 3 and 1 reductions are shown in the top panels, and the same, but with the {\tt Fitsky}  = 2 reduction, are shown in the bottom panels.
We divided the mosaic field into the inner ($SMA\leq 14\arcmin$, left panels) and outer ($SMA> 14\arcmin$, right panels) regions to investigate if there is any variation dependent on the stellar crowding.

We found that the three sky fitting options output very similar photometry;
in all the cases, the magnitude offsets are almost negligible at the bright side ($F814W\lesssim24$~mag), and they are only on the order of 0.01~mag at the anticipated level of the TRGB ($F814W\approx25.4$~mag).
Slight differences can be seen in the milli-magnitude level such that 
the {\tt Fitsky} = 1 reduction appears to provide a better agreement than the 
{\tt Fitsky} = 2 reduction, as the offsets are $\sim$0.005~mag smaller.
Similarly, the offsets are smaller in the inner region by $\sim$0.015~mag than the outer regions.
We note, however, that both of the {\tt Fitsky} = 1 and 2 reductions are fainter in all  cases than the {\tt Fitsky} = 3 reduction.
The photometric accuracy of these reductions can be tested with artificial stars and we provide further details in \autoref{app:appa}.
We found that our main photometry dataset with {\tt Fitsky} = 3 is more accurate than the other two reductions ({\tt Fitsky} = 1 and 2), but there is a slight systematic offset; the recovered magnitudes are fainter than their intrinsic values with $\Delta F814W = 0.01$~mag at $F814W \sim25.4$~mag (see \autoref{fig:fake_fitsky}).

\subsection{PSF radius}\label{sec:psfsize}

\begin{figure}
    \centering
    \includegraphics[width=0.45\textwidth]{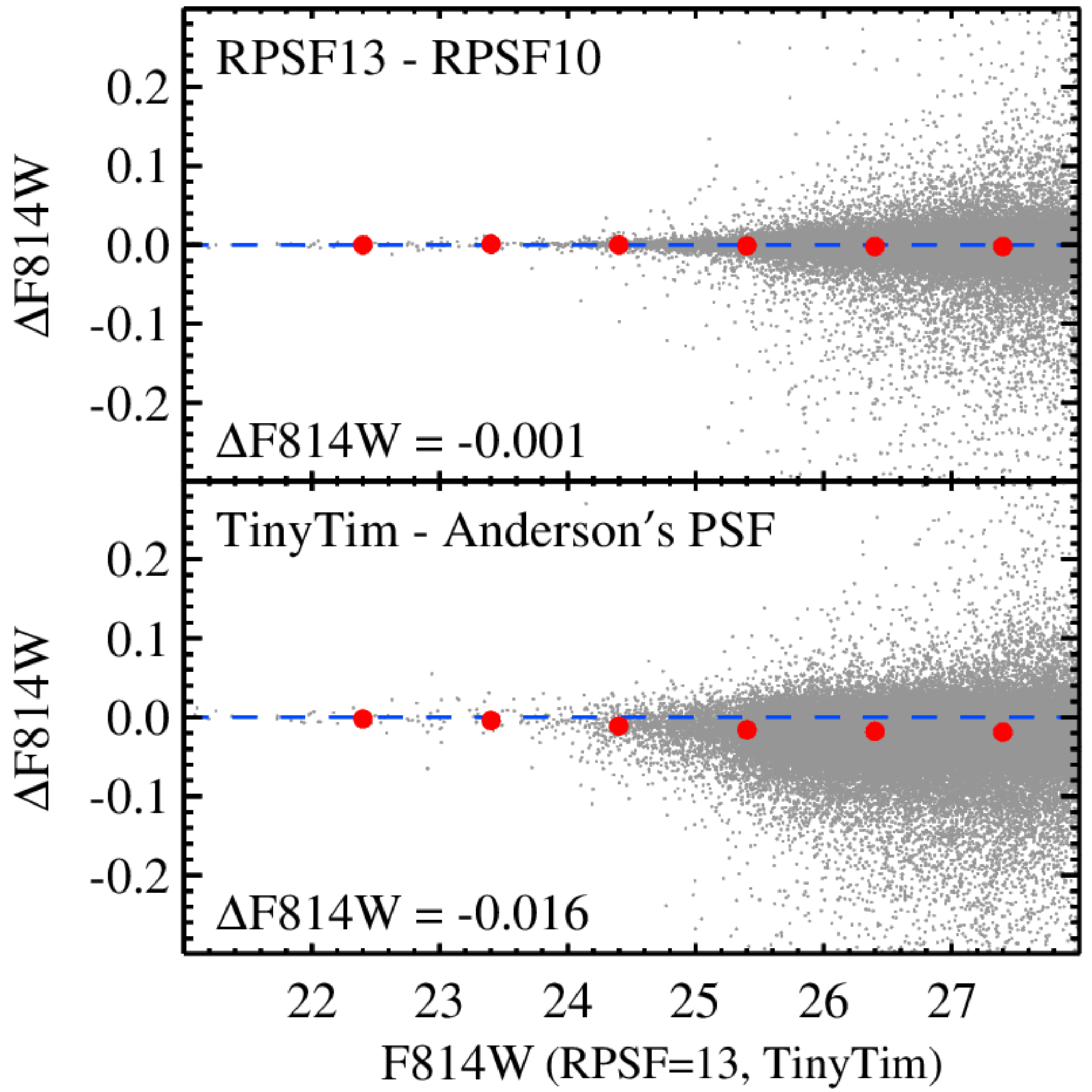}
    \caption{
    Comparison of PSF fitting choices.
    The top panel shows the impact of different PSF radii ({\tt RPSF} = 10 and 13 pixels).
    The bottom panel shows the impact of different PSF models: TinyTim PSF \citep{Krist_2011} and ``Anderson's PSF'' \citep[e.g.,][]{Anderson_2006}. 
     Median offsets in each magnitude bin are indicated by red dots.
    The median offsets at the anticipated magnitude of the TRGB are also indicated in the figure.
    }
    \label{fig:rpsf}
\end{figure}

The parameter {\tt RPSF} determines the size of the PSF radius used for star subtraction.
The PSF radius should be sufficiently larger than the full-width at half maximum of the PSF, and thus becomes more important in crowded-field photometry.
The DOLPHOT manual recommends the {\tt RPSF} values either of 10 or 13~pixels (depending on the version of the manual; v1.1 to v2.0, respectively); here we test our photometry with both values as is shown in the top panel of \autoref{fig:rpsf}. 
We confirmed that the choice of the {\tt RPSF} values, either 10 or 13~pixels, does not meaningfully change the output magnitudes.
The median offset at $F814W\sim25.4$~mag, the vicinity of the TRGB, is only 1~milli-magnitude.
The spatial selection does not change the results; both inner ($SMA\leq14\arcmin$) and outer ($SMA>14\arcmin$) regions show the same offset, $\Delta F814W = -0.001$~mag.

\subsection{TinyTim versus Anderson's PSFs}\label{sec:psftype}

There are two types of PSFs available within DOLPHOT: TinyTim PSFs \citep{Krist_2011} and Jay Anderson's PSFs \citep[e.g.,][]{Anderson_2006}.
While the current version of DOLPHOT does not recommend using the Jay Anderson PSFs due to issues with the implementation of PSF libraries in DOLPHOT, we have tested to see the effects of adopting different PSFs.
We found that the choice of PSF types has only a minimal impact; the median difference is negligible at the bright side ($F814W\lesssim24$~mag) and only $\Delta F814W = - 0.016$ mag at the expected magnitude of the TRGB (\autoref{fig:rpsf}, bottom).
The degree of difference is almost the same in the region interior ($\Delta F814W = -0.016$~mag) and exterior ($\Delta F814W = -0.017$~mag) to SMA = 14$'$.

Here we emphasize that all our reductions presented in this paper and all of the reductions used for the CCHP \citep[e.g.,][and references therein]{Freedman_2019} have been carried out using a PSF model determined from the TinyTim PSFs.
As a result, PSF-dependent variations are minimized within the CCHP work, but this is a systematic to consider when adopting literature measurements.

\subsection{Summary}\label{sec:phot_valid_summary}

We summarize here the wide range of  stringent photometry comparisons undertaken in this section: 
different photometry packages (DAOPHOT vs. DOLPHOT), 
different image types (FLC vs. DRC frames),
different local sky fitting methods, and 
different PSF models (synthetic Tiny-Tim, Anderson's Core, 
and empirically measured PSFs). 
Furthermore, these techniques have also been tested for different levels of stellar crowding (i.e., inner vs. outer regions).
From the full range of tests explored in this section, 
we conclude that an additional systematic uncertainty of 0.02 mag can be added to our final photometry used to measure the TRGB. 
This conservative estimate of the uncertainty comes mostly from the choice of local sky fitting method, though additional tests explored in \autoref{app:appa} have shown that our choice of {\tt Fitsky=3} is the most reliable. 
We included this additional systematic uncertainty in our final error budget in \autoref{sec:geocalib}. 

\begin{figure*}
    \centering
    \includegraphics[width=0.9\textwidth]{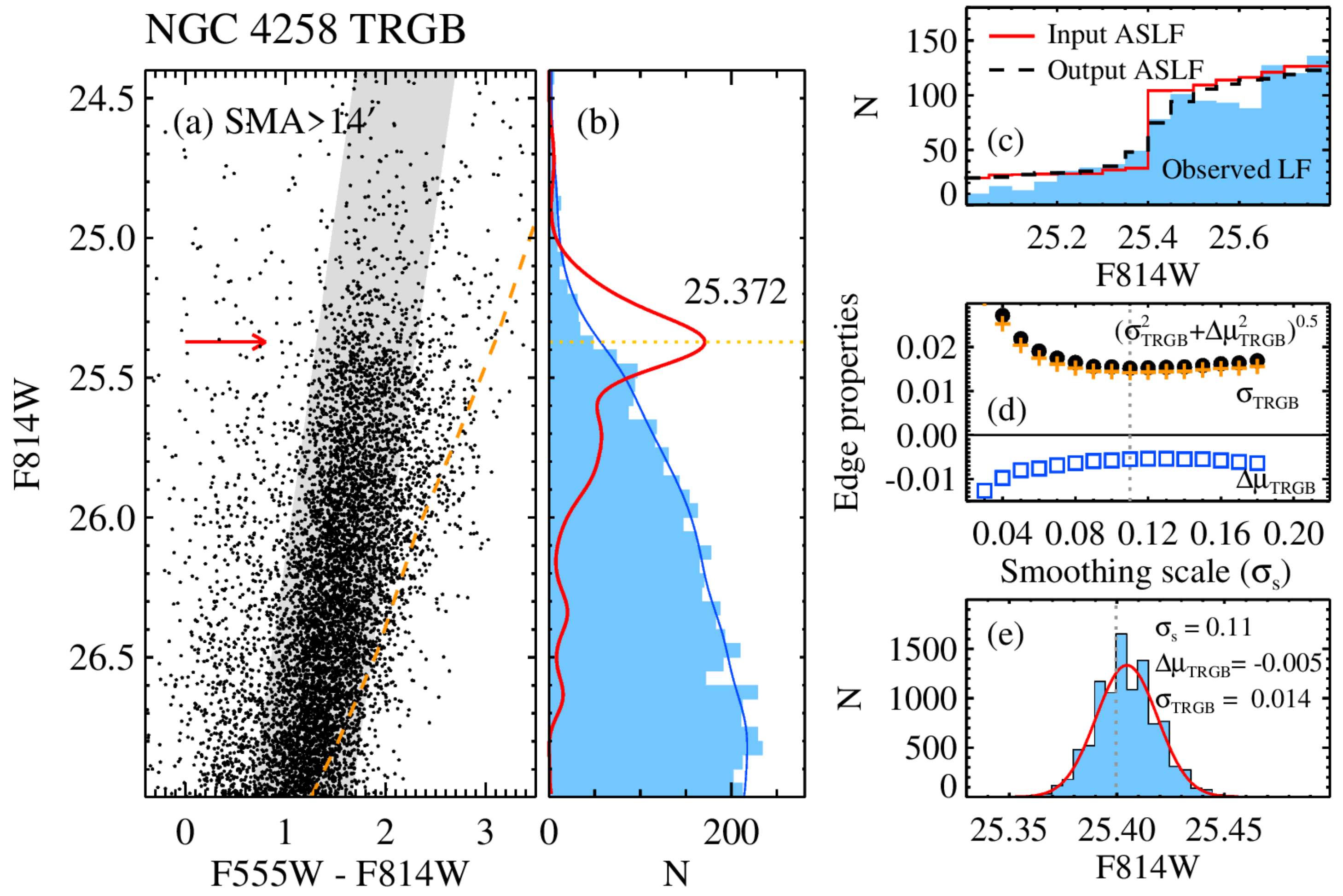}
    \caption{
    The CCHP determination of the apparent magnitude of the TRGB in \gal. We use only sources more than $SMA=14\arcmin$ from the center of \gal, as shown in \autoref{fig:imagedata}. (a) CMD of resolved stars beyond $14\arcmin$. There are approximately 3000 stars between F814W = 25.4 mag and 26.4 mag. 
    A yellow dashed line indicates the 50\% completeness level determined from the artificial star tests.
    (b) Luminosity function (LF) binned at 0.05 mag intervals (blue histogram), also shown as a smoothed curve (blue). Both the LF and the Sobel-edge response (red) were smoothed with \sigmasmooth=\sigmasmoothval~mag. 
    The smoothed luminosity function is shown by a thin blue line in the middle panel.
    There is a clear and unambiguous peak in the Sobel edge-detector response at F814W = \trgbobsvalROUNDED~mag. 
    (c) Determination of statistical and systematic uncertainties using artificial luminosity functions inserted (red) and recovered (black) from the images. The luminosity function of real stars is shown by a filled histogram for comparison.
    (d) and (e) Optimization of the smoothing scale and estimation of errors.
    A quadratic sum of the statistical ($\sigma_{\rm TRGB}$) and systematic ($\Delta\mu_{\rm TRGB}$) uncertainties is observed to be minimized when the smoothing scale is set to \sigmasmooth= \sigmasmoothval~mag. 
    Our analysis suggests $\sigma_{\rm stat} = \sigma_{\rm TRGB} = \pm \trgbobsvalstaterr$~mag and $\sigma_{\rm sys} = \Delta\mu_{\rm TRGB} = \pm \trgbobsvalsyserr$~mag.}
    \label{fig:ourtrgb}
\end{figure*}

\section{Measuring the TRGB in NGC~4258}\label{sec:trgb}

In this section, we present our measurement of the apparent magnitude of the TRGB in the uncrowded, dust-free halo of  \gal. 
Throughout this process, we make decisions that will be explicitly justified in the sections to follow, and that align with those adopted by the CCHP in prior papers. 
 
We included only those stars with a semi-major axis, galactocentric distance larger than $14\arcmin$ (31 kpc). 
At this extent the mean surface brightness is below $\mu_{\rm B} \sim 28$~mag arcsec$^{-2}$ \citep{Watkins_2016}. 
The choice to photometer stars in the low-surface-brightness outskirts has several advantages in making a clean determination 
of the TRGB magnitude.
Outside the adopted boundary, the colors of the majority of RGB stars are consistent with them being metal-poor, thus minimizing any trend of the ($I$-band) TRGB absolute magnitude with color.
The RGB stars are also spatially well separated, and the systematic impact from crowding and/or a varying sky background due to unresolved light is thereby minimized. 
As a result, we see less variation of the observed TRGB magnitude, which indicates that these lines of sight are not passing {\it through} any extended gas disk of \gal, suggesting that the effect of in situ extinction is minimized as well. 

The process for measuring the apparent magnitude of the TRGB is described in 
\autoref{fig:ourtrgb}.
We follow the general procedures that were established by \citet{Hatt_2017}, and refined in later CCHP works.
We use the region exterior to SMA = 14$'$ shown in the upper right panel of \autoref{fig:imagedata}; even with this restriction, a sample of approximately 3,000 stars remains over the F814W one-magnitude range from 25.4 to 26.4~mag. \autoref{fig:ourtrgb}(a) and \autoref{fig:ourtrgb}(b) show the CMD (black points), the luminosity function (blue), and edge-detection (red) peak response for this selection.
The luminosity function (LF) is derived from the stars in the shaded region of the CMD, which has a slope of $\Delta\rm{F814W}/\Delta\rm{(F555W-F814W)}=-2.5$~mag with a color boundary of $\rm{(F555W-F814W)}=$ 1.3 and 2.3~mag at the anticipated magnitude of the TRGB.
This color-magnitude selection ensures that those sources inconsistent with being blue (metal-poor) RGB stars are not contaminating the marginalized LF.
We varied the blue edge of the color cut from $\rm{(F555W-F814W)}= 1.0$~mag to $\rm{(F555W-F814W)}= 1.3$~mag and found that the effect on our TRGB detection was less than 0.01 mag.
The color-selected LF is finely binned at  0.001 mag, and then smoothed using the GLOESS algorithm with a Gaussian smoothing scale of \sigmasmooth= $\pm$\sigmasmoothval~mag. 
A signal-to-noise weighted Sobel kernel [-1, 0, 1] is applied to the smoothed LF to determine the magnitude where the LF has its greatest change.
We detect the apparent magnitude of the TRGB at F814W = $\trgbobsval$~mag.

Our uncertainties are determined as shown in the right three panels of \autoref{fig:ourtrgb}. 
We used a sample of artificial stars that satisfy the color and magnitude boundaries used for the TRGB detection given above. 
An idealized input luminosity function for the RGB+AGB is generated from artificial stars and its recovered luminosity function is then used for the TRGB detection (\autoref{fig:ourtrgb}(c)).
Because we know the input TRGB precisely, we can measure any offset from the input value thereby providing a systematic uncertainty, with the scatter about the mean being indicative of the statistical uncertainty.
Following \citet{Hatt_2017}, a series of tests on our artificial luminosity functions have shown that the total TRGB detection uncertainty can be minimized when a Gaussian smoothing scale of \sigmasmooth= $\pm$\sigmasmoothval~mag is applied (\autoref{fig:ourtrgb}(d)).
At this smoothing scale, the statistical and systematic errors are estimated to be $\sigma_{\rm stat}$ = $\pm$\trgbobsvalstaterr~mag and $\sigma_{\rm sys}$ = $\pm$\trgbobsvalsyserr~mag (\autoref{fig:ourtrgb}(e)).
We tested a range of reasonable smoothing scales (large enough to smooth over Poisson noise, but not too large that the edge is systematically displaced) and find that the effect on the measured TRGB magnitude is 0.01 mag or less.

In summary, for \gal\ we have measured the TRGB apparent magnitude for RGB stars in a large region with SMA \textgreater\ 14$'$, giving $F814W_{\rm TRGB}$ = \trgbobsval~mag with $\sigma_{\rm stat}$ = $\pm$\trgbobsvalstaterr~mag and $\sigma_{\rm sys}$ = $\pm$\trgbobsvalsyserr~mag.

\begin{figure*}
    \centering
    \includegraphics[width=0.9\textwidth]{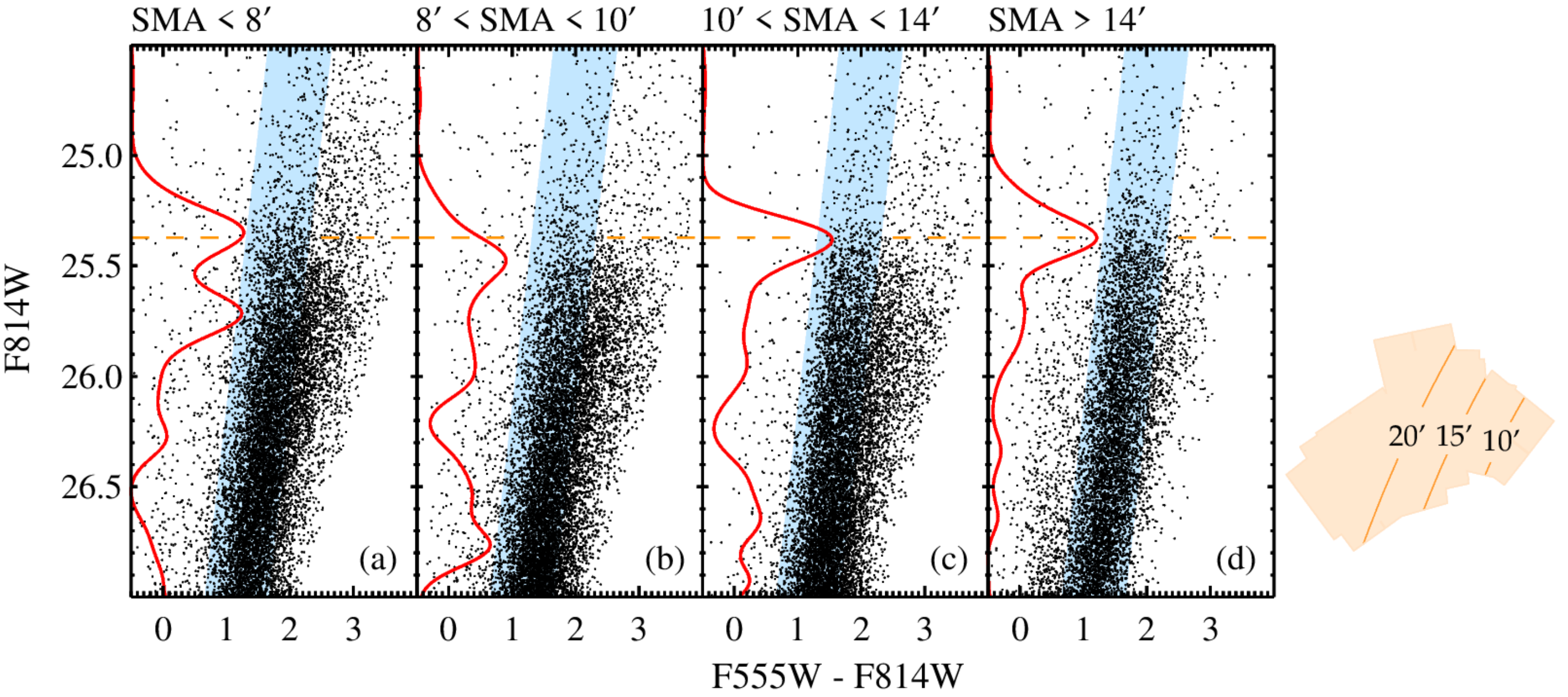}
    \caption{
    CMDs of resolved stars in four regions of the \gal mosaic field: $SMA\leq8\arcmin$ (a),  $8\arcmin < SMA \leq 10\arcmin$ (b),  $10\arcmin < SMA \leq 14\arcmin$ (c), and $SMA > 14\arcmin$ (d).
    We selected the blue RGB stars in the shaded region of each CMD and used them for the edge detection (red line).
    Our optimum detection of the TRGB ($F814W_{\rm TRGB}$ = \trgbobsval~mag) taken from the outer region with $SMA>14\arcmin$ (d) is shown by a dashed line in all panels.
    A schematic view of the mosaic field is shown on the right.
    }
    \label{fig:cmd_4fields}
\end{figure*}

 \subsection{Variation of the TRGB in the mosaic field}\label{sec:minor}
 
In \autoref{app:appb}, we discuss the details of optimizing our spatial selection of old and blue (metal-poor) RGB stars.
\autoref{fig:cmd_4fields} displays CMDs of the mosaic field. 
The spatial selection is the same as in \autoref{fig:cmd_5fields}, such that the number of stars in the blue RGB domain (shaded region) is approximately the same from field to field, thereby minimizing sample-size systematics.
The edge-detection algorithm applied to the blue RGB stars finds maximum responses at (a) $F814W = 25.346 \pm 0.020$~mag, (b) $25.476\pm 0.017$~mag, (c) $25.383\pm 0.014$~mag, and (d) $25.372\pm0.016$~mag.
Here the errors are the quadratic sum of the statistical and systematic uncertainties measured from a series of tests with artificial-star luminosity functions.

It is found that the outer regions with $SMA > 10\arcmin$ (\autoref{fig:cmd_4fields}(c) and \autoref{fig:cmd_4fields}(d)) have a consistent and stable measurement of the TRGB. 
There is one prominent peak in the edge detection response at the expected magnitude of the TRGB, where the discontinuity in the luminosity function of the stars in CMDs is clearly seen.
The measured TRGB magnitudes are almost identical; 
indeed the difference between the TRGB detections measured in the outer two fields (0.011~mag) is smaller than their individually quoted $1\sigma$ errors (0.015~mag).
For the inner two regions with $SMA < 10\arcmin$ (\autoref{fig:cmd_4fields}(a) and \autoref{fig:cmd_4fields}(b)), we begin to detect slight variations, such that the edge detection responses are fainter (by nearly 0.1~mag) for \autoref{fig:cmd_4fields}(b), or ambiguous, with two equally-significant peaks seen in \autoref{fig:cmd_4fields}(a).
We suspect that the origin of the unstable measurement is due to the inclusion of disk stars with a possible spread in age and metallicity, together with dust extinction in the plane of the disk, and that these different effects all play a role (see \autoref{app:appb}). 
 It is well known that for these reasons, the TRGB method is most precise and accurate when applied to stars in the halos of galaxies, and not in their disks.

\begin{figure}
    \centering
      \includegraphics[width=0.45\textwidth]{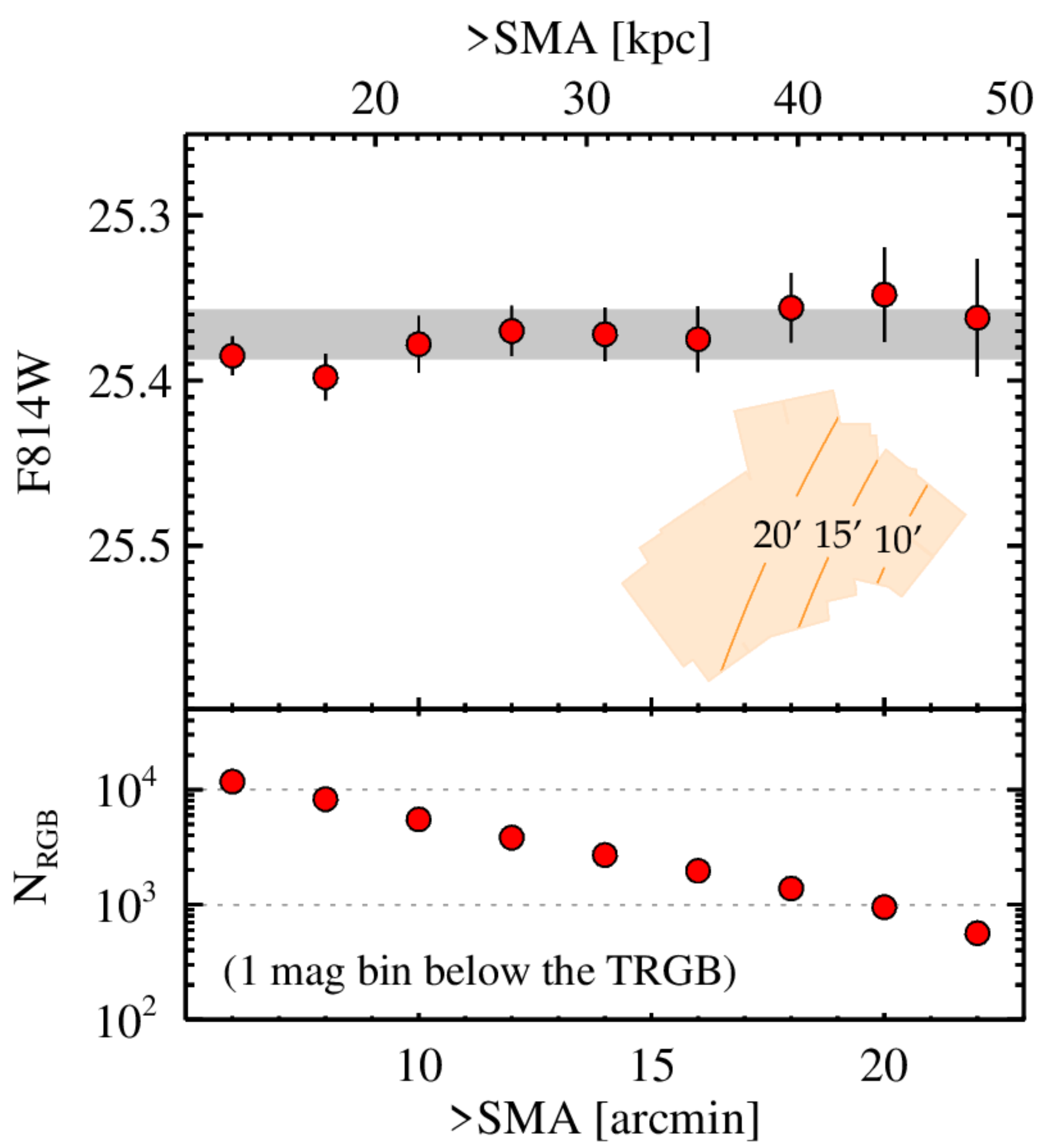}
    \caption{
    Top Panel: Variation of the measured TRGB magnitude as a function of the SMA.  
    Stars beyond the given radial distance in SMA are used for the TRGB detection.
    Also shown, inset in the upper panel, is a schematic view of the footprint of the mosaic field with boundaries of three representative radial distances: $SMA = 10\arcmin, 15\arcmin,$ and $20\arcmin$.
    Bottom Panel: Number of stars in the one-magnitude bin below the TRGB as a function of SMA. 
    }
    \label{fig:sma_trgb}
\end{figure}

We next consider optimizing the detection of the TRGB in the mosaic field.
\autoref{fig:sma_trgb} shows the variation of the TRGB magnitudes as a function of the spatial selection. 
The blue RGB stars beyond the given radial distance are selected and used for the TRGB detection (upper panel).
The measurement errors become larger for the outer regions because of the diminishing numbers of stars available to define the TRGB (bottom panel).
Nevertheless, the measured TRGB magnitudes appear to be largely consistent over the entire mosaic, but we caution that those measurements including the innermost regions ($SMA\lesssim 10\arcmin$) should be considered less reliable due to the unstable nature of the TRGB detections at this radius, as shown in \autoref{fig:cmd_4fields}.
Similarly, the measurements relying only on the outermost regions ($SMA\gtrsim18\arcmin$) have less weight because the lower number of stars involved in the detection,  which could induce a bias in the TRGB detection \citep{MadoreFreedman_1995, Madore_2009}, if the luminosity function is not being completely filled at the brightest magnitudes.
It is not obvious how best to impose a spatial cut for the optimal selection of the TRGB, given that each of the measurements agree well within their $1\sigma$ uncertainty, but we estimate that a region with $SMA\gtrsim14\arcmin$ is an optimal choice.
Nevertheless, we can see that our detection of the TRGB is very insensitive to the spatial selection, and therefore the systematic uncertainty associated with this choice is negligibly small.

\begin{figure*}
    \centering
   \includegraphics[width=0.7\textwidth]{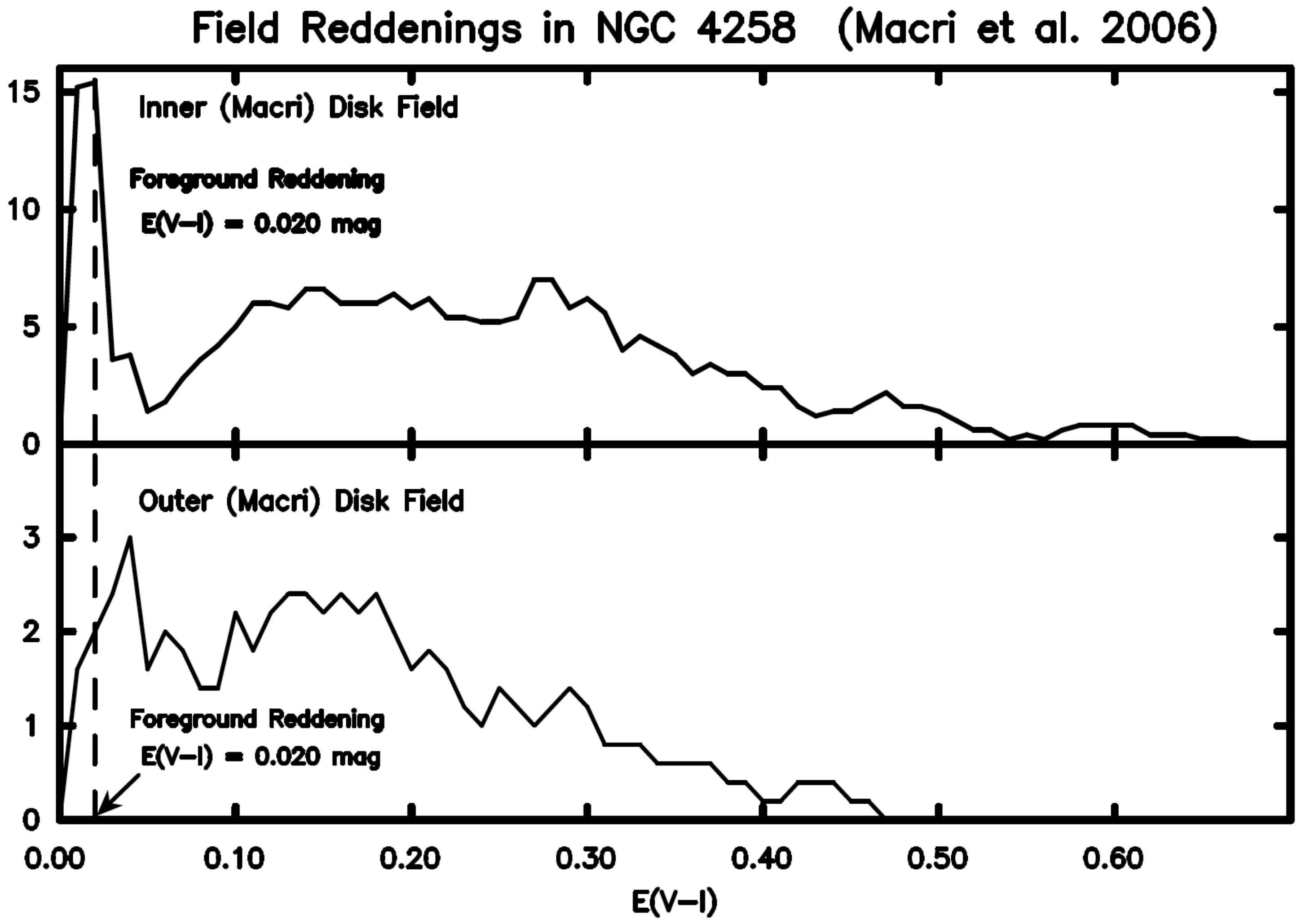}
    \caption{Smoothed histograms of individual Cepheid $E(V-I)$ reddenings in \gal, as published by Macri et al. (2006) for their ``Inner Disk" field (upper panel) and their ``Outer Disk" field (our Disk field; lower panel). The Milky Way foreground reddening is shown by the dashed vertical line at $E(V-I) = $ 0.020~mag. Both fields have considerable reddening: the Inner field has a median reddening of about 0.25~mag, while the Outer field has a median reddening of 0.15~mag in E(V-I). }
    \label{fig:ceph_red}
    \end{figure*}

\subsection{The TRGB in the ``Disk'' Field: A Cautionary Tale} \label{sec:outerdisk}

In this section, we turn our attention to the measurement of TRGB stars in the high-surface-brightness ``Disk" region of \gal. In particular, we illustrate why an unbiased TRGB measurement cannot be made in this field, due to a confluence of reddening, population, and crowding effects. This deep ACS field lies in the southern disk of \gal, shown by a blue square in  \autoref{fig:imagedata}.
The pointing was chosen and the observations designed with
the intent of discovering of Cepheid variables \citep[][PID = 9810]{hst_prop_9810}. As such, multiple visits were required to discover the variables and determine their periods, mean magnitudes and colors.
Total exposure times were 18,400s in F555W and 9,200s in F814W-band; significantly longer than the median exposure time of the mosaic field (2780s in F814W), for example.

The field spans from $SMA = 6\arcmin$ to $10\farcm5$, wherein the $B$-band surface brightness ranges from $\mu_B \simeq 24$ to $26$~mag arcsec$^{-2}$ \citep{Watkins_2016}, approximately $3\sim4$ mag arcsec$^{-2}$ brighter than the outer region we used for the optimal TRGB detection (see also \autoref{fig:rdp_3rgbs}). 
In addition to their determination of the maser distance to NGC 4258, \citet{Reid_2019} undertook  a calibration of the TRGB absolute magnitude based on a TRGB measurement in this same Disk 
field presented in \citet{Macri_2006} (which they call ``Outer Disk''). For their calibration, they chose a value of zero for reddening intrinsic to the host galaxy at this position in the disk. More recently, \citet{Nataf_2020} make the same assumption in their calibration of the Cepheid Period-Color relation using this field.

We point out that the assumption of zero internal reddening along the line of sight to this Disk field is directly contradicted by Cepheid results previously presented in \citet{Macri_2006}, who use the exact same Disk dataset. Their uncorrected TRGB magnitude is $I = 25.42 \pm 0.02 $ mag. However, through their multi-wavelength analysis of the Cepheid Leavitt Law, these authors found a differential modulus $ \mu_V - \mu_I = 0.14 \pm 0.06 $ mag.  This observed difference in the $V$ and $I$ distance moduli provides direct evidence for some degree of reddening to RGB stars projected into this field.

Indeed, the Cepheid analysis, given in \citet{Macri_2006}, makes it quite clear that these stars are significantly reddened: they actually measure individual line-of-sight reddenings on a star-by-star basis (as given in their {Table 6}).
In \autoref{fig:ceph_red} we show histograms of their reddenings determined for Cepheids in both their Inner and Outer disk fields. There is no question that the Cepheids in these fields are reddened and extincted. While the Population II RGB stars are expected to have lower extinction than Cepheids, given the non-zero reddening confirmed across the disk field, it is hard to rule out the possibility that RGB stars projected into the Disk field will not also suffer from extinction. We view the Cepheid reddening values both as an upper limit and as evidence for non-zero reddening to RGB stars in this Disk field.

\begin{figure*}
    \centering
    \includegraphics[width=0.9\textwidth]{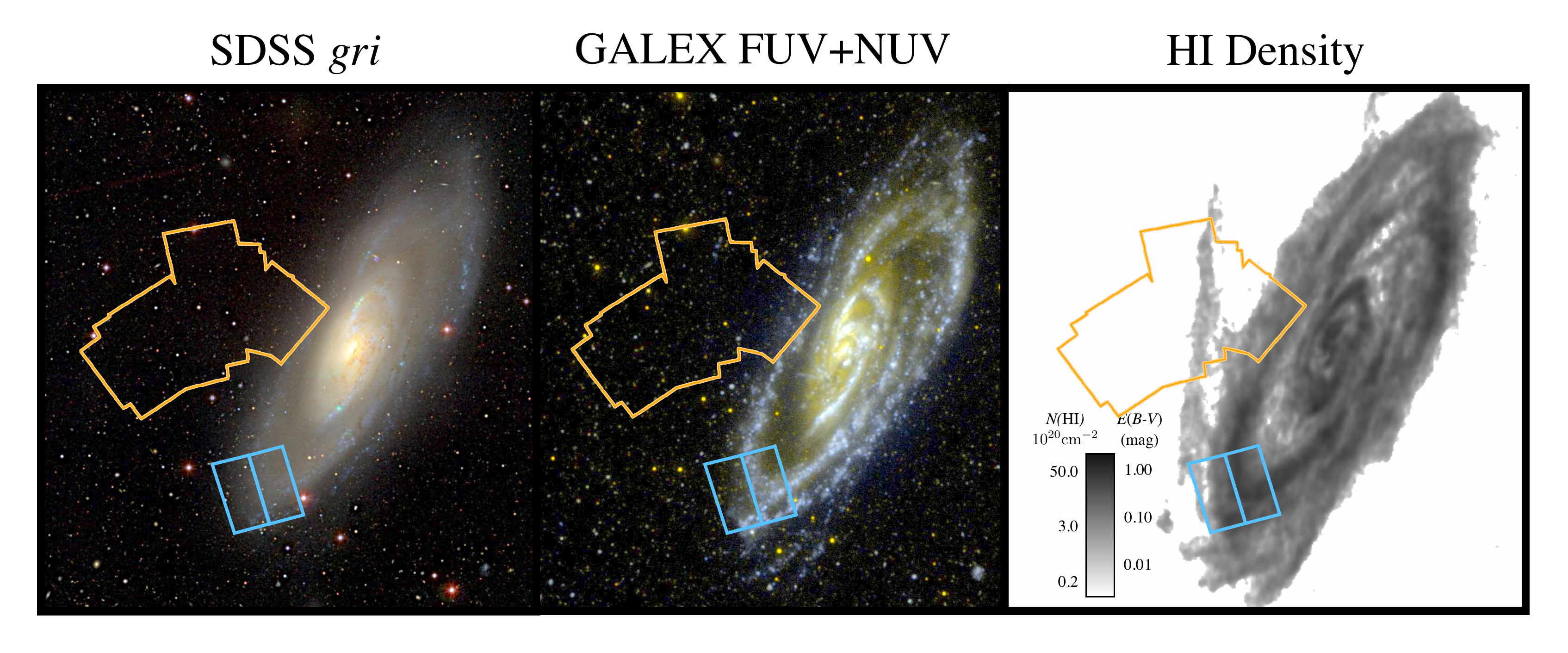}
    \caption{Multi-wavelength diagnostics for \gal: SDSS $gri$ color image (left), Galex FUV+NUV color image (middle), and HI Map from \citet{Heald_2011} (right). The HI column densities are used to estimate values of $E(B-V)$ based on conversions provided in \citet{Bohlin_1978}. The halo mosaic 
    field (orange) and Disk field (blue) are overplotted for reference.
     }
    \label{fig:multiwave}
\end{figure*}

Adding further evidence in support of the fact that the Disk field is problematic for a measurement of the TRGB, in \autoref{fig:multiwave} we present multi-wavelength imaging across the body of \gal: SDSS $gri$ (left), GALEX FUV+NUV (middle) and neutral hydrogen column density \citep[right,][]{Heald_2011}. 
The $gri$ imaging suggests the presence of blue star-forming regions in the Disk field, which is strengthened by the existence of UV-bright counterparts to these suspected star-forming regions. Both panels strongly suggest that there is a significant population of young and intermediate-aged stars distributed throughout this disk field. Together with the HI column density measurements, we display an example mapping between column density to color excess, from \citet{Bohlin_1978}. Significant star formation is indicated by the shorter wavelength imaging, in addition to evidence from the ensemble of multi-wavelength data that there exist nonzero quantities of internal dust extinction in and across the lines of sight to this field.

\begin{figure*}
    \centering
   \includegraphics[width=0.9\textwidth]{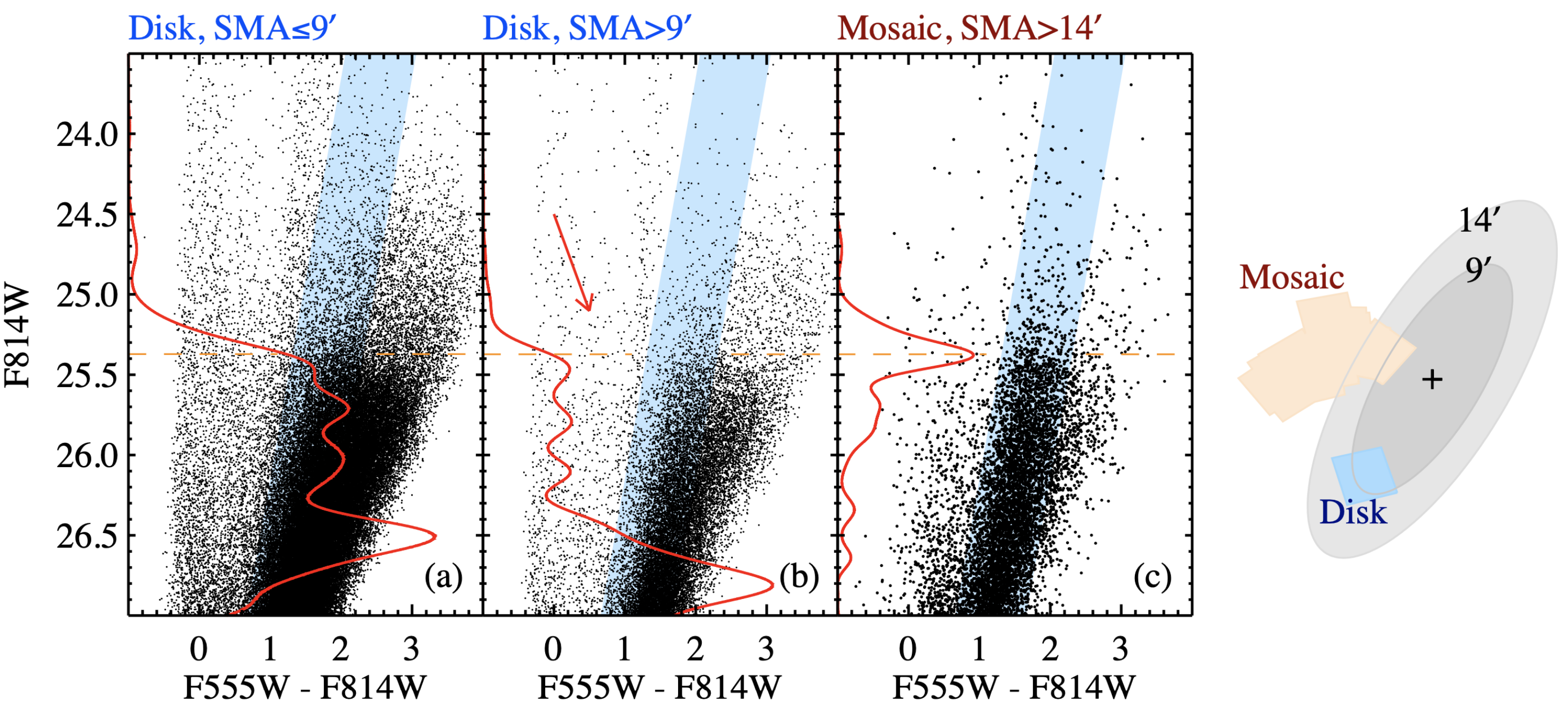}
    \caption{
    CMDs of the inner ($SMA\leq9\arcmin$) and outer($SMA>9\arcmin$) regions of the disk field. 
    The third panel shows our CMD of the outer region of the mosaic field ($SMA\geq14\arcmin$), which was selected for the optimum determination of the \gal TRGB.
    As in our earlier analysis, we used the blue RGB stars in the shaded regions of the CMDs for the edge detection (red line). An arrow marked in the middle panel (b) shows a reddening vector of $A_I = 0.6$~mag.
    A schematic view of \gal and the (colored) footprints of the two fields discussed in this section, are shown to the far right, outside of the panels.
    }
    \label{fig:cmd_maj_min}
\end{figure*}

To further explore the effects of young stellar populations and dust in the Disk field, we turn to comparisons of resolved stellar photometry and edge detection response functions between the Disk field and our adopted halo mosaic selection.
In \autoref{fig:cmd_maj_min} we display CMDs of the inner ($SMA\leq9\arcmin$ of the blue region in the right-most panel) and outer ($SMA>9\arcmin$ of the blue region in the right-most panel) regions of the Disk field (\autoref{fig:cmd_maj_min}(a) and \autoref{fig:cmd_maj_min}(b)).
For comparison, \autoref{fig:cmd_maj_min}(c) is a CMD of the halo dominated region ($SMA>14\arcmin$ (the orange region shown in the right-most panel) we used for the optimal detection of the TRGB.
Overplotted in red on each CMD is an edge detection response function, computed from stars contained within the blue color-selection region, introduced in the previous section. For reference, the CMD and response function shown in panel (c) are equivalent to those shown in \autoref{fig:ourtrgb} and \autoref{fig:cmd_4fields}(d). The yellow-dashed line demarcates the TRGB magnitude $F814W = \trgbobsvalROUNDED$ mag, as measured from our adopted mosaic halo dataset.

The inner region of the Disk field (\autoref{fig:cmd_maj_min}a) exhibits clear signs of young and intermediate-aged stellar populations: a Population I (blue) main sequence (the vertical feature at $F555W - F814W \approx 0$) and 
cooler helium-burning, red supergiants  
(the slanted feature reaching to F814W = 23.5~mag at $F555W - F814W \approx 1.6$).
These populations should not be surprising, given that this disk field was chosen to sample Cepheid variables that are themselves slightly evolved but still young Population I supergiants. A sharp truncation in source counts, which could be seen as evidence for the TRGB, can be seen near $F814W \simeq 25.5$~mag and $F555W-F814W = 2.2$, located almost entirely outside of our color selection region.

A small number of stars belonging to the main sequence and He-burning sequences identified in \autoref{fig:cmd_maj_min}(a), are visible in \autoref{fig:cmd_maj_min}(b) in the outer portion of the Disk field ($SMA>9\arcmin$; see inset). Still, the locus of the RGB is displaced to significantly redder colors than those shown in the halo region of panel (c), again lying almost entirely outside of our color selection. In the absence of a significant population of young and intermediate-aged stars, the remaining explanation would appear to be a combination of metallicity and reddening effects. 
For reference, a reddening vector with magnitude $A_{F814W} = 0.6$ mag is shown.

The CMD presented in panel (c) of \autoref{fig:cmd_maj_min} shows those sources located in the outermost region of the mosaic dataset, which we adopted as our halo selection. Comparing to the two Disk CMDs, it is immediately apparent  that the RGB stars present in either of the Disk CMDs in panels (a) and (b) do not belong to an old, unreddened, or metal-poor population, as required to make an unbiased measurement of the TRGB. 

How might the observations described in this section affect empirical measurements of the TRGB in the Disk field? To answer that question, we examine the edge detection response functions, overplotted in red in the CMDs in \autoref{fig:cmd_maj_min}. We first notice that, in stark contrast to the strongly peaked edge detection response we saw in our adopted halo mosaic dataset, reproduced for reference in panel (c), there is no single, clear peak in the response functions shown for the two Disk regions. This multi-peaked structure in the response function could be the result of several entangled effects: mixed stellar populations, differential extinction, or source crowding. The observable impact of this ambiguous, multi-peaked degeneracy is borne out in the literature: \citet{Macri_2006} found a TRGB magnitude $I = 25.42 \pm 0.02 $ mag for the Disk field; and in the same Disk field \citet{Rizzi_2007} measured a TRGB magnitude of $I = 25.52$ mag. 
The two differ by 0.1 mag, which is perhaps a better reflection of the uncertainties associated with measuring the TRGB in this Disk dataset. 

To test whether our conclusions drawn from the CMDs could be caused by non-physical, systematic offsets in, for example, photometric colors across independent datasets, we repeated our DAOPHOT-DOLPHOT comparison for the Disk field, in a manner identical to that described in Section 3.
A star-by-star comparison was carried out, from which we found that the two independent catalogs of stellar photometry agree well. The mean offset is measured to be smaller than $\Delta F814W = 0.01$~mag.

In summary, the tests described above demonstrate that the Disk region is problematic for measurement of the TRGB. First, the multi-wavelength Leavitt Law provides a constraint on the line-of-sight reddening to individual Cepheids \citep[e.g.,][]{freedman_1988,Freedman2001}. We show the results of this analysis in \autoref{fig:ceph_red} for Cepheids in the Disk field, providing an upper limit and evidence for non-zero reddening to RGB stars located in the same field. Secondly, the maps presented in \autoref{fig:multiwave} provide observational evidence of interstellar gas/dust, and recent star formation in the Disk region, suggesting the Disk field suffers from population-mixing effects and reddening. Third, from the CMDs in \autoref{fig:cmd_maj_min}, we showed that there is a clear redward displacement of the RGB locus in the Disk region with respect to that in the Halo selection, potentially suggestive of metallicity or reddening effects, or both. With a wide range of unconstrained astrophysics at play, we conclude that the TRGB cannot be accurately measured in the \gal\ Disk field, and that our measurement, which uses the mosaiced halo dataset, provides the current best, unbiased measurement of the Population II TRGB in \gal.

\begin{deluxetable*}{llllll} 
\tabletypesize{\small}
\setlength{\tabcolsep}{0.05in}
\tablecaption{Previous TRGB Measurements \label{tab:prev_trgb} }
\tablewidth{0pt}
\tablehead{\colhead{Study} & \colhead{Camera} & \colhead{Field} & \colhead{Technique} & \colhead{System} & \colhead{TRGB Mag}} 
\startdata
\citet{Mouhcine_2005}    &  WFPC2 & near F1\tablenotemark{a}       & ML    & $I$   & 25.25$^{+0.13}_{-0.02}$\\
                         &        &                             & Sobel & $I$   & 25.22 $\pm$ 0.09 \\
&&&&&\\
\citet{Macri_2006}       &  ACS   & Outer Disk\tablenotemark{b} & Sobel & $I$ & 25.42 $\pm$ 0.02 \\ 
&&&&&\\
\citet{Rizzi_2007}       &  ACS   & Outer Disk\tablenotemark{b} & ML    & $I$ & 25.52 \\ 
                         &  ACS   & Inner Disk\tablenotemark{b} & ML    & $I$ & 25.45 \\
                         &        &                             & Mean  & $I$ & 25.49 $\pm$ 0.06 \\
&&&&&\\
\citet{Mager_2008}       &  ACS   & F1\tablenotemark{c}         & Sobel & $T$\tablenotemark{d}   & 25.24 $\pm$ 0.04 \\
                         &  WFPC2 & near F1\tablenotemark{a}       & Sobel & $T$\tablenotemark{d}   & 25.20 $\pm$ 0.06 \\
&&&&&\\
\citet{Madore_2009}      &  ACS   & F1\tablenotemark{c}         & Sobel & $I$   & 25.39 $\pm$ 0.11 \\
                         &        &                             & Sobel & $T$\tablenotemark{d}   & 25.21 \\
&&&&&\\
\citet{JangLee_2017}     &  ACS   & F1\tablenotemark{c}         & Sobel & F814W & 25.382$\pm$ 0.031\\
                         &        &                             & Sobel & $QT$\tablenotemark{d}  & 25.370$\pm$ 0.023\\
&&&&&\\
\citet{Jacobs_2009}  &  ACS   & F1\tablenotemark{c}         & ML    & F814W & 25.43 $\pm$ 0.03 \\
     &     &         &     & $I$ & 25.46 $\pm$ 0.03 \\
&&&&&\\
\hline
This Work                &  ACS   &Mosaic\tablenotemark{c,e}   & Sobel & F814W &  \trgbobsvalROUNDED $\pm$ \trgbobsvalstaterrROUNDED $\pm$ \trgbobsvalsyserrROUNDED \\
\enddata
\tablenotetext{a}{\hst\ Program GO-9086, \citet{hst_prop_9086}}
\tablenotetext{b}{\hst\ Program GO-9810, \citet{hst_prop_9810}}
\tablenotetext{c}{\hst\ Program GO-9477, \citet{hst_prop_9477}}
\tablenotetext{d}{ Corrected for the foreground extinction ($A_{\rm I} \sim A_{\rm F814W} = 0.025$ mag)}
\tablenotetext{e}{\hst\ Program Go-10399 \citet{hst_prop_10399}}
\end{deluxetable*} 

\section{Comparison With Other TRGB Measurements} \label{sec:trgbcompare}

In the following, we summarize the previous TRGB measurements in \gal with a tabulation presented in \autoref{tab:prev_trgb}. 
We first describe each of the studies individually, and close by discussing them in aggregate.

    In an early study, \citet{Mouhcine_2005} reduced a WFPC2 pointing on the minor axis \citep[][PID = 9086]{hst_prop_9086}, which is similar in location to field F1, used in this study.
    \citeauthor{Mouhcine_2005} performed aperture photometry to extract the magnitudes and colors of resolved stars. \citeauthor{Mouhcine_2005} then obtained two measurements of the TRGB, one using a maximum-likelihood based technique and the other using a Sobel edge detector, 
    finding $I_{\rm TRGB}=25.25^{+0.13}_{-0.02}$~mag for the former and $I_{\rm TRGB}=25.22\pm0.09$~mag for the latter.
    Further discussion of this field is given in \autoref{app:appc}.

    Subsequently \citet{Macri_2006} used an $ACS$ pointing into the disk of \gal\ situated on the major axis at SMA $\sim 8\arcmin$ \citep[][PID = 9810]{hst_prop_9810}, which they call the ``Outer Disk" field (our Disk field).
    This pointing was specifically designed for the discovery of Population I Cepheid variables.
    Using a Sobel edge detector, they found $I_{\rm TRGB}=25.42\pm0.02$~mag. 
    
    \citet{Rizzi_2007} reduced three $ACS$ pointings around \gal: one on the minor axis \citep[][PID = 9477]{hst_prop_9477}, and two on the major axis, which are the ``Inner Disk" and ``Outer Disk" fields from \citet{Macri_2006}.
    The minor axis field was used for the color-dependent calibration of the TRGB.
    The two disk fields were used to compare their TRGB distance to \gal with that from Cepheids in \citet{Macri_2006}. 
    \citeauthor{Rizzi_2007} measured the TRGB using a Maximum-Likelihood 
    (ML) 
    luminosity function fitting technique \citep[developed by][]{Makarov_2006} and found $I_{\rm TRGB}$ = 25.45 and 25.52 mag for the \citeauthor{Macri_2006} ``Inner Disk" and ``Outer Disk" regions, respectively; they take the average of the two distances for a final result of $I_{\rm TRGB}$=25.49$\pm$0.05~mag.

    \citet{Mager_2008} processed two fields: one taken with ACS \citep[][PID = 9477]{hst_prop_9477} and a field at a similar position with WFPC2 \citep[][PID = 9086]{hst_prop_9086}.
    Both fields were on the minor axis with the intention of avoiding the disk of \gal. 
    They introduced a color-corrected tip magnitude for the TRGB detection defined by \citet{Madore_2009} as 
    $T = I_0 - 0.20[(V-I)_0-1.5]$.
    Using a Sobel edge detector, run on the $T$-band luminosity function they found statistically consistent results from the two fields: $T_{\rm RGB} = 25.24 \pm 0.04$ mag from the ACS and $25.20 \pm 0.06$ mag from the WFPC2.

    \citet{Madore_2009} used the same data processing of the ACS field as in \citeauthor{Mager_2008} and determined the TRGB in both $I$ and $T$ systems.
    The reported values are: $I_{\rm TRGB}$ = 25.39 $\pm$0.11~mag and $T_{\rm RGB}$ = 25.21 mag, the latter being similar to that of \citet{Mager_2008} in the $T$ system.
    Though, we note the color-coefficients on the $T$ system were different.
    
    Most recently, \citet{JangLee_2017} used the ACS field on the minor axis \citep[][PID = 9477]{hst_prop_9477}, which is identical to the ACS field studied in \citet{Mager_2008}, \citet{Madore_2009}, and named F1 in the present paper.
    They provided two estimates of the TRGB, one in the native F814W system using the stars in the blue color bin ($1.0 \lesssim (F555W - F814W)_0 \lesssim 2.1$), and another one 
    in a modified photometric system similar to $T$, but using a quadratic form for the TRGB slope, called $QT$\footnote{$QT =F814W_{\rm 0} - 0.116(Color - 1.6)^2 + 0.043(Color - 1.6)$, where $Color = (F555W - F814W)_0$}, applied to all the stars without any additional, color-based selections.
    Their uncorrected (apparent) values for the TRGB magnitude are $F814W_{TRGB}= 25.382$~mag, and $QT_{\rm RGB}= 25.395$~mag.
    The resulting values (corrected for reddening), adopting a Milky Way extinction of $A_{\rm F814W} = 0.025$~mag \citep{Schlafly_2011}, are  $F814W_{\rm 0,TRGB} = 25.357 \pm 0.031$~mag, and $QT_{\rm RGB}= 25.370 \pm 0.023$~mag.
   
    We note that the Extra-galactic Distance Database (EDD) \citep{Jacobs_2009} has a large number of TRGB-based distances, based on homogeneously processed photometry and uniformly-derived TRGB detections determined with the \citet{Makarov_2006} ML fitting method. 
    The team applies the TRGB zero-point calibration of \citet{Rizzi_2007} that includes a metallicity correction computed as a function of the mean color of the TRGB; the EDD reports each step in the distance determination. 
    The EDD tip detection for our F1 ACS pointing \citep[][PID = 9477]{hst_prop_9477} is $F814W_{\rm TRGB}$ = 25.43~$\pm$~0.03~mag at a mean RGB color of $(F555W-F814W) = 2.12^{+0.04}_{-0.03}$. 
    We note that their mean color is close to the red-edge of our blue RGB selection box.

As can be seen from inspection of \autoref{tab:prev_trgb}, the TRGB measurements are on four photometric systems: $I$, $T$, F814W, and $QT$, and involve three main fields: a ``Disk'' ACS field, the inner Halo ACS field (our F1), and the WFPC2 halo field near F1. 
The different photometric systems used in each study make it difficult 
to see a true study-to-study variation
of the TRGB magnitudes. We therefore selected those measurements in the native $F814W$ or $I$ system with modern ACS data (disk field or F1) and compared them with our outer halo based measurement.
We found that these literature TRGB measurements cover a total range of 0.14~mag (25.38 -- 25.52~mag) with a mean of 25.43~mag and a standard deviation of 0.05~mag.
Our TRGB measurement (\trgbobsvalROUNDED $\pm$ \trgbobsvalstaterrROUNDED $\pm$ \trgbobsvalsyserrROUNDED~mag) is, therefore, in agreement with most of the other literature TRGB values within the mutual uncertainties.


\section{Geometric Calibration of the TRGB Zero Point} \label{sec:geocalib}

\gal is a known host to an $\rm H_2O$ megamaser, which arises from stimulated emission from compact molecular clouds induced by hard radiation from the supermassive black hole about which they orbit.
Its proximity ($\sim 7$ Mpc) enables high-precision measurement of the proper motions and radial velocities of each maser using very long baseline interferometric (VLBI) mapping techniques.
The orbits of the masers can be well approximated by a Keplerian rotation curve, resulting in an accurate distance to the galaxy \citep{Greenhill_1995,Reid_2019}.
Our measurement of the \gal TRGB can be combined with this geometric distance, to provide an absolute calibration of the TRGB luminosity.

We start from our optimal estimation of the \gal TRGB, $F814W_{\rm TRGB} = \trgbobsval \pm \trgbobsvalstaterr$~(statistical) $\pm \trgbobsvalsyserr$~(systematic)~mag, which is based on the blue (metal-poor) RGB stars in the halo regions  ($SMA \gtrsim 30$~kpc), where the in situ extinction should be negligible. 
As in our previous CCHP papers, we consider systematic uncertainties associated with the photometric calibration: \ZPerr~mag for the ACS photometric zero-point,  \EEerr~mag for the encircled energy curves for stars cooler than $K$-type, and \Apcorrerr~mag for the aperture correction.
In addition to these systematic uncertainties related to the absolute photometric calibration, our detailed technical investigation of the photometry codes (\autoref{sec:validation}) and TRGB detection process (see \autoref{sec:trgb}) has revealed the following additional systematic terms: 
  a term for the choice of photometry code and PSF modeling of \photchoiceerr~mag, 
  a term for the color selection of \trgbcolselerr~mag , and 
  a term for effects from selection of a single LF smoothing scale of \trgbsmoothselerr~mag. 

The Milky Way foreground extinction toward \gal is known to be small, $A_{F814W} = \IextinctioROUNDED \pm \IextinctionerrTOTAL$~mag \citep{Schlafly_2011}.
Here the error is taken from a quadratic sum of half of the extinction ($\sigma_A=0.013$~mag), and an additional systematic uncertainty ($\sigma_{halo}=0.01$~mag) based on the large scale statistical analysis of halo reddenings undertaken by \citet{Peek_2015}. 
Finally, for \gal we adopt the most recently published  geometric distance and its associated errors, those being \maserdistmodwerr \citep{Reid_2019}. 

With all of these terms taken into account, we obtain a TRGB zero-point of \trgbzeropoint with a total error of $\pm \trgbzerototalerr$~mag (2.3\% in distance).
A summary of our revised TRGB zero-point error budget is given in \autoref{tab:our_trgb_calib}.

\begin{deluxetable}{lccc}[h] 
\tabletypesize{\footnotesize}
\setlength{\tabcolsep}{0.05in}
\tablecaption{Error Budget for \gal\ TRGB Calibration \label{tab:our_trgb_calib}}
\tablewidth{0pt}
\tablehead{ 
\colhead{Sources of Uncertainty for}  &  \colhead{Final} & \colhead{$\sigma_{stat}$} & \colhead{$\sigma_{sys}$} \\
\colhead{TRGB Apparent Distance Modulus}  &  \colhead{Values} & \colhead{(mag)} & \colhead{(mag)}
}
\startdata
\multicolumn{4}{l}{}\\
Edge Detection&       & ~~\trgbobsvalstaterr & ~~\trgbobsvalsyserr \\
Photometry Choices           &       &       \nodata       & \photchoiceerr \\
Color Selection              &       &       \nodata       & \trgbcolselerr \\
Smoothing Selection         &       &        \nodata       & \trgbsmoothselerr \\
STScI ACS F814W ZP           &       &       \nodata        &  \ZPerr \\
STScI ACS F814W EE Correction &       &      \nodata       &  \EEerr \\
Empirical Aperture Correction      &       &       \nodata      & \Apcorrerr \\ 
\hline \hline
TRGB$_{\rm F814W}$ (apparent) & \trgbobsvalROUNDED  & \trgbobsvalstaterrROUNDED & \trgbobsvalsyserrROUNDED \\ 
$A_{\mathrm{F814W}}$ (Galactic foreground)	      & ~\IextinctioROUNDED  & \nodata                   & ~~\IextinctionerrTOTAL \tablenotemark{a} \\
TRGB$_{F814W}$ (true)	      & 25.347  & \nodata                   & \nodata \\
\hline  
NGC~4258 Maser Distance Modulus\tablenotemark{b}& \maserdistmod       & \maserdistmodstaterr      & \maserdistmodsyserr \\
\hline \hline 
$M_{F814W}^{\rm TRGB }$   & \trgbzero  &  \trgbzerostaterr & \trgbzerosyserr
\enddata
\tablenotetext{a}{Taken to be half of $A_{\mathrm{F814W}}$ and including a 0.01~mag component from internal extinction.}
\tablenotetext{b}{\citet{Reid_2019}}
\end{deluxetable} 

\subsection{Discussion}

    The calibration of the TRGB given in this paper,  \trgbzeropoint, calibrated by the geometric megamaser distance to \gal, is in excellent agreement with the totally independent calibration of the TRGB from  \citet{Freedman_2020} using the DEB distance to the LMC, which gives $M_{F814W}^{TRGB} = $ -4.054 $\pm$ 0.022 (stat) $\pm$ 0.039 (sys).

    For completeness, we simply note that our measurement also meets the additional cross-checks against other TRGB calibrations in the Local Group provided in \citeauthor{Freedman_2019}; specifically, $M_{I}^{TRGB} = -4.09\pm0.03 (\rm stat) \pm0.05 (\rm sys)$ based on a DEB geometric distance  to the SMC, and $M_{I}^{TRGB} = -4.056\pm0.053 (\rm stat) \pm0.080 (\rm sys)$ based on a composite sample of Galactic globular clusters covering a range of metallicities.
    
    Our calibration is also within one sigma of the value reported by \citet{Reid_2019}, that being $M_{F814W}^{TRGB} = -4.01 \pm 0.04$~mag.
    Because we used the same maser distance estimate to \gal as determined by and used in \citet{Reid_2019}, it can be shown that the +0.04~mag difference in the reported zero-points comes directly from the TRGB magnitude differences: $F814W_0 = 25.385 \pm 0.030$~mag in \citeauthor{Reid_2019}, 
    and $F814W_0 = \trgbredcorrval~\pm~\trgbobsvalstaterr$~mag this study.
    The TRGB magnitude adopted by \citet{Reid_2019} was, in turn, derived from two previously published studies: (a) \citet{Macri_2006}, which is based on the Disk field (where $F814W_0=25.398\pm0.02$~mag), and (b) \citet{JangLee_2017} who used Field~1, the innermost portion of the Halo mosaic field (where $F814W_0=25.357\pm0.031$~mag) (see \autoref{fig:imagedata} for the field configurations).
    These two measurements are systematically +0.05 and +0.01~mag fainter, respectively, than our TRGB measurement, where the differences can be attributed to a combination of astrophysical systematics present in the Disk field (see Section 4.2).

    Our TRGB measurement is fully independent of that described in \citet{JangLee_2017}: 
     (1) We have analyzed the region exterior to SMA=14$'$ not included in their study. 
     (2) We have undertaken a fully independent processing/reduction of the photometry. And 
     (3) We have utilized an independent CCHP edge-detection strategy. 
    We arrive at statistically identical results for our two (present and past) TRGB measurements ($F814W_0 = \trgbredcorrval \pm \trgbobsvalstaterr$~mag in this study and $F814W_0 = 25.357\pm0.031$~mag in \citet{JangLee_2017}).
    For a color selection of $(F555W-F814W) < 2.1$~mag, \citeauthor{JangLee_2017} find a ``Blue-TRGB'',  absolute magnitude of $M^{\rm TRGB}_{\rm F814W} = -4.030 \pm 0.068$~mag (their Table 6). 
    This value is based on the distance to NGC 4258 in \citet{Riess_2016} ($\mu_0 = 29.387 \pm 0.049 \pm 0.029$~mag).
    Updating this distance to the \citet{Reid_2019} value, the \citeauthor{JangLee_2017} TRGB zero-point becomes $M^{\rm TRGB}_{\rm F814W} = -4.041 \pm 0.049$~mag, which is also consistent with the value in this paper.
    \citet{JangLee_2017} also provide a zero-point of the TRGB based on the LMC as an anchor, giving $M_{F814W}^{TRGB} = -3.96 \pm 0.11$~mag.
    There are three sources that contribute to their relatively large uncertainty:
    (a) the TRGB detection ($\sigma = 0.042$~mag), 
    (b) the $I$-band extinction ($\sigma = 0.07$~mag), and 
    (c) the uncertainty, at that time, in the distance to the LMC ($\sigma = 0.049$~mag).
    While this zero-point is +0.087 mag fainter than our new determination based on \gal, its large error results in a low-level ($0.7\sigma$) statistical difference.
    
    Our determination of the absolute magnitude of the $I$-band TRGB is -0.081 mag ($1.2\sigma$) brighter than a recent determination by \citet{Yuan_2019} ($M_{F814W}^{TRGB} = -3.97 \pm 0.046$~mag). 
    There are four elements that are used to construct the \citeauthor{Yuan_2019} zero-point: 
    1) the apparent magnitude of the LMC TRGB \citep[taken from][]{JangLee_2017}, 
    2) an accounting for the stellar crowding of the OGLE photometry \citep{Yuan_2019}, 
    3) the $I$-band extinction towards the LMC \citep[taken from][]{ Haschke_2011}, and 
    4) the eclipsing binary distance to the LMC \citep[taken from][]{Pietrzynski_2019}.
    \citeauthor{Yuan_2019} have shown that the crowding-dependent bias in the OGLE $I$-band photometry is very small, being on the order of 0.01~mag, even in the disk region.
    They obtained the LMC TRGB magnitude from \citet{JangLee_2017}, who provided the TRGB magnitudes in several regions of the LMC.
    After applying their corrections for stellar crowding and the filter transformation (from $I$ to F814W), \citeauthor{Yuan_2019} derived a mean TRGB magnitude for the LMC, $F814W_0 = 14.507 \pm 0.012$~(stat) $\pm 0.036$~(sys) mag.
    Here the statistical error is derived from the standard deviation of the eight TRGB values divided by the square root of the degrees of freedom. 
   However, the locations of the eight TRGB regions measured by \citet{JangLee_2017} have considerable overlap (see Fig 9 of \citet{JangLee_2017}).  
    The TRGB measurements from these regions are thus not statistically independent, which may lead to an underestimate of the uncertainty, if not taken into account.

    The systematic error of the LMC TRGB in \citeauthor{Yuan_2019} ($\pm0.036$~mag) is dominated by the uncertainty they adopted for the line-of-sight extinction ($\pm0.03$~mag).
    These authors used the extinction map provided by \citet{Haschke_2011}, who measured the reddening using the mean color of the red clump stars. 
    While the red clump has been used as a reddening indicator, the intrinsic color of the LMC red clump is not currently well constrained, ranging from $(V-I)_0 = 0.84$ to 0.93~mag \citep{Haschke_2011, Gorski_2020, Nataf_2020, Skowron_2020}.
 This would suggest that techniques using the red clump to measure reddenings may be subject to a sizable systematic uncertainty.
Taking into account these additional uncertainties, and the conclusions drawn in \citet{Freedman_2020}, we conclude that the \citeauthor{Yuan_2019} zero-point agrees with ours at the one-sigma level.
    
    Our determination of the $I$-band TRGB zero-point, $M_{F814W}^{TRGB} =$ \trgbzero~mag, is consistent with
    the canonical value of $M_{F814W} \sim M_I = -4.05$~mag \citep[][]{Rizzi_2007, Bellazzini_2008, Tammann_2008, Madore_2009}; however, our estimated error is now about 50\% smaller than previous TRGB calibration errors.
    
\section{Summary} \label{sec:summary}

The main objective of this paper has been to establish a highly accurate and precise geometric calibration of the TRGB method directly in the HST F814W  ``flight magnitude" system. The 15 TRGB calibration fields in \gal are located in the outer, gas and dust-free halo of the galaxy \gal, which has a measured geometric distance based measurements from the 22 GHz water masers in the accretion disk around its central black hole. 

We have undertaken independent DAOPHOT and DOLPHOT analyses, quantifying the uncertainties in our point source photometry. We present a robust detection of the TRGB at $F814W = ~$\trgbobsvalROUNDED $\pm$ \trgbobsvalstaterrROUNDED~(stat)~mag based on over 3,000 stars (within one magnitude of the TRGB) having mean photometric errors of $\pm$0.06 mag. 
Applying a Milky Way foreground extinction correction of $A_{F814W} = $ \IextinctioROUNDED~mag, and subtracting the maser distance modulus gives $M_{F814W}^{TRGB} = $ \trgbzero $\pm$ \trgbzerostaterr (stat) $\pm$ \trgbzerosyserr (sys)~mag. 

Our new, direct-to-HST calibration is completely independent of, and agrees well with, an earlier calibration of \citep{Freedman_2019,Freedman_2020}, based on a recent geometric DEB distance to the LMC \citep{Pietrzynski_2019}. This study resulted in a TRGB calibration of $M_{F814W}^{TRGB} = $ -4.054 $\pm$ 0.022 (stat) $\pm$ 0.039 (sys)~mag. The LMC-based calibration, however, relies on transformations from \citet{Riess_2016} to convert observations in ground-based $I$ to the F814W system. The new calibration presented here bypasses those transformations, and their uncertainties, altogether.

With internal consistency established, we can simply average the two independent CCHP calibrations to determine an updated TRGB zero point of $M_{F814W}^{TRGB} = -4.053$~mag, where the total uncertainty is now reduced by $\sim$30\% to $\pm0.034$~mag (or 1.6\% in distance). The broader implication, the impact of adding this second geometric anchor to the TRGB distance scale, means that the conclusions reached in \citet{Freedman_2019, Freedman_2020} are virtually unchanged in their magnitude, while being significantly strengthened in their statistical and systematic certainty.

\acknowledgments


We thank Peter Stetson for a copy of the \textsc{daophot} family of programs as well as his helpful interactions on its use for our science case. I.S.J. is grateful to Dr. Roelof. de Jong for helpful discussions on data reduction and stellar populations in nearby galaxies over the past few years.

Support for program \#13691 (PI W.L Freedman) was provided by NASA through a grant from the Space Telescope Science Institute, which is operated by the Association of Universities for Research in Astronomy, Inc., under NASA contract NAS 5-26555.
Support for this work was also provided by NASA through Hubble Fellowship grant \#51386.01 awarded to R.L.B. by the Space Telescope Science Institute, which is operated by the Association of Universities for Research in Astronomy, Inc., for NASA, under contract NAS 5-26555. M.G.L. is supported by a grant from
the National Research Foundation (NRF) of
Korea, funded by the Korean Government (NRF-2019R1A2C2084019).

Some of the data presented in this paper were obtained from the Mikulski Archive for Space Telescopes (MAST). STScI is operated by the Association of Universities for Research in Astronomy, Inc., under NASA contract NAS 5-26555. 

This research has made use of the NASA/IPAC Extragalactic Database (NED) and the NASA/IPAC Infrared Science Archive (IRSA), both of which are operated by the Jet Propulsion Laboratory, California Institute of Technology, under contract with the National Aeronautics and Space Administration.
We also acknowledge the use of the HyperLeda database (\url{http://leda.univ-lyon1.fr}).






Funding for the Sloan Digital Sky Survey IV has been provided by the Alfred P. Sloan Foundation, the U.S. Department of Energy Office of Science, and the Participating Institutions. SDSS-IV acknowledges support and resources from the Center for High-Performance Computing at the University of Utah. The SDSS web site is \url{www.sdss.org}.

SDSS-IV is managed by the Astrophysical Research Consortium for the Participating Institutions of the SDSS Collaboration including the Brazilian Participation Group, the Carnegie Institution for Science, Carnegie Mellon University, the Chilean Participation Group, the French Participation Group, Harvard-Smithsonian Center for Astrophysics, Instituto de Astrof\'isica de Canarias, The Johns Hopkins University, Kavli Institute for the Physics and Mathematics of the Universe (IPMU) / University of Tokyo, the Korean Participation Group, Lawrence Berkeley National Laboratory, Leibniz Institut f\"ur Astrophysik Potsdam (AIP),  Max-Planck-Institut f\"ur Astronomie (MPIA Heidelberg),  Max-Planck-Institut f\"ur Astrophysik (MPA Garching),  Max-Planck-Institut f\"ur Extraterrestrische Physik (MPE), National Astronomical Observatories of China, New Mexico State University, New York University, University of Notre Dame,  Observat\'ario Nacional / MCTI, The Ohio State University, Pennsylvania State University, Shanghai Astronomical Observatory, United Kingdom Participation Group,Universidad Nacional Aut\'onoma de M\'exico, University of Arizona, University of Colorado Boulder, University of Oxford, University of Portsmouth, University of Utah, University of Virginia, University of Washington, University of Wisconsin,  Vanderbilt University, and Yale University.

Some of the figures used in this paper were based on observations made with the NASA Galaxy Evolution Explorer (GALEX), which was operated for NASA by the California Institute of Technology under NASA contract NAS 5-98034.  

Finally, we thank the University of Chicago and the Carnegie Institution for Science for their long-term support of our continuing investigation into the expansion rate of the Universe.

\vspace{5mm}
\facilities{HST(ACS/WFC)}

\software{
          Astropy \citep{2013A&A...558A..33A, 2018AJ....156..123A},
          DAOPHOT \citep{stetson_1987},
          DOLPHOT \citep{2016ascl.soft08013D},
          Drizzlepac \citep{2012ascl.soft12011S, 2013ASPC..475...49H, 2015ASPC..495..281A},
          Matplotlib \citep{2007CSE.....9...90H},
          NumPy \citep{numpy},
          SciPy \citep{scipy},
          }
\bibliography{cchp_papers,ho,ngc4258,submitted_papers,all_other_papers}{} 
\bibliographystyle{aasjournal}

\appendix

\section{photometric accuracy with different sky fitting methods} \label{app:appa}

\begin{figure}
\figurenum{A1}
    \centering
    \includegraphics[width=0.7\textwidth]{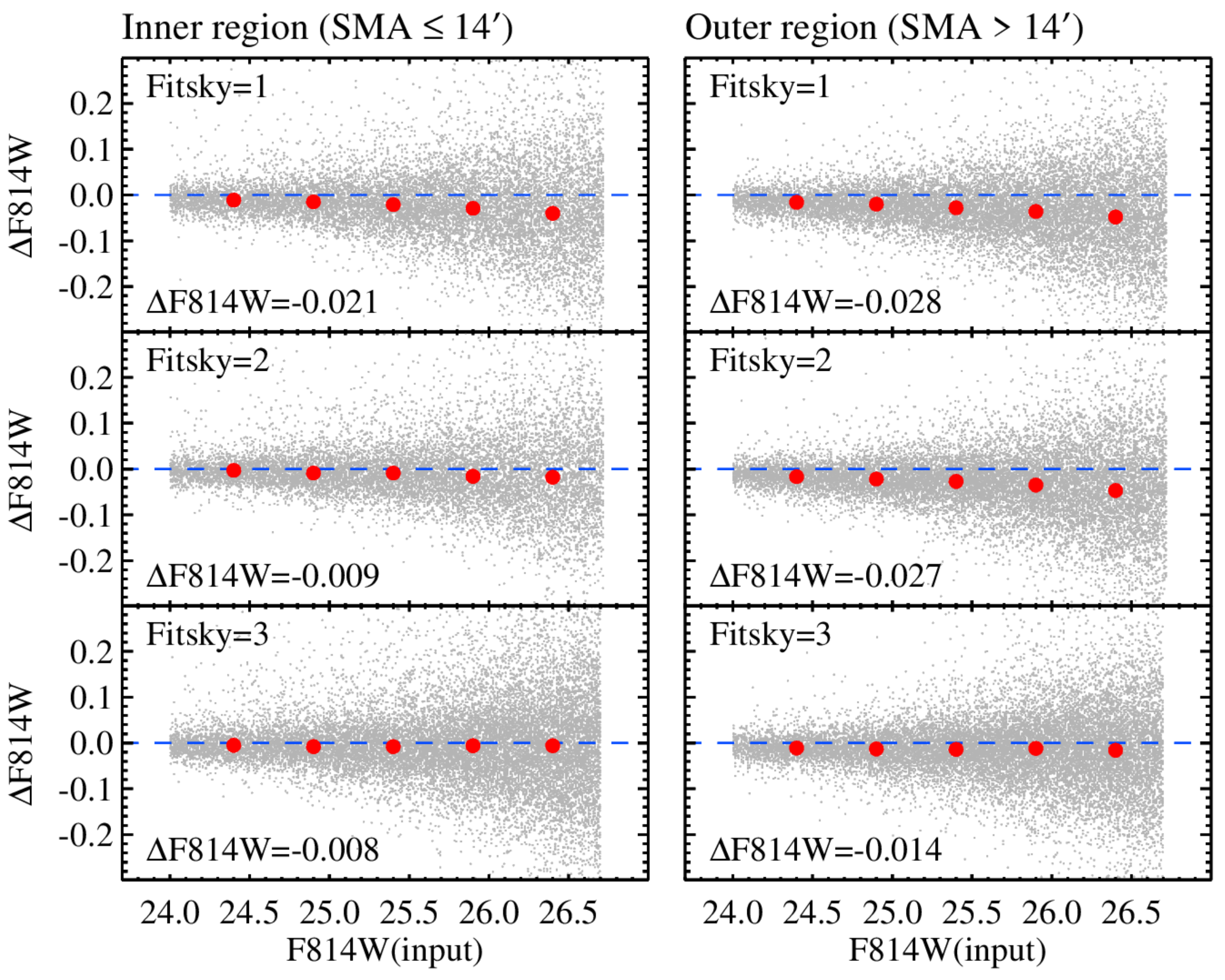}
    \caption{Differences in F814W (inputs minus outputs) vs. input F814W magnitudes for the inner (left) and outer regions (right) with different sky fitting options: {\tt Fitsky} = 1 (top), 2 (middle), and 3 (bottom). 
    The median offset at the TRGB level ($F814W\sim25.4$~mag) is marked in each panel.
    }
    \label{fig:fake_fitsky}
\end{figure}

\autoref{fig:fake_fitsky} displays the results of artificial star tests with the three different sky fitting options ({\tt Fitsky}~= 1, 2, and 3) in DOLPHOT.
Panels are split into two groups, one for the inner regions ($SMA\leq14\arcmin$, left) and the other one for the outer regions ($SMA>14\arcmin$, right).
The input artificial stars have a mean color of (F555W - F814W) $\sim$ 1.5, similar to the sequence of the blue RGB stars in the \gal halo.
Red dots show median offsets in each mangitude bin and the offset at the TRGB level is marked in each panel.
The error of the median offset is dominated by the statistical error that is estimated to be $\sigma_{\Delta F814W} \simeq0.002$~mag. 

We confirm that the median offsets do not exceed 0.03~mag in all the cases, and they are much smaller than the mean photometric error of $\pm0.09$~mag at the TRGB level.
The offsets are all negative, indicating that the recovered magnitudes are fainter (but only slightly by 0.01 $\sim$ 0.03~mag) than their intrinsic values.
A relative difference is noted such that the {\tt Fitsky = 3} reduction shows a better agreement with smaller offsets than the other two options ({\tt Fitsky = 1} and 2).
This is consistent with the results from the real star photometry shown in \autoref{fig:fitsky_real}; there we found that the {\tt Fitsky = 1} and 2 reductions are slightly fainter by $0.01 \sim 0.02$~mag than the {\tt Fitsky = 3} reduction.

This test shows that all the three sky fitting options in DOLPHOT output reliable photometry, but the accuracy can be enhanced when the {\tt Fitsky = 3} option is used in this dataset. 
We remind the reader that our main photometry dataset was reduced with  {\tt Fitsky = 3}, so that our measurement of the TRGB is based upon  the most reliable photometry.

\section{Optimizing Detection of the Old, Blue Halo Population For TRGB Measurements} \label{app:appb}

\begin{figure}
\figurenum{A2}
    \centering
    \includegraphics[width=0.9\textwidth]{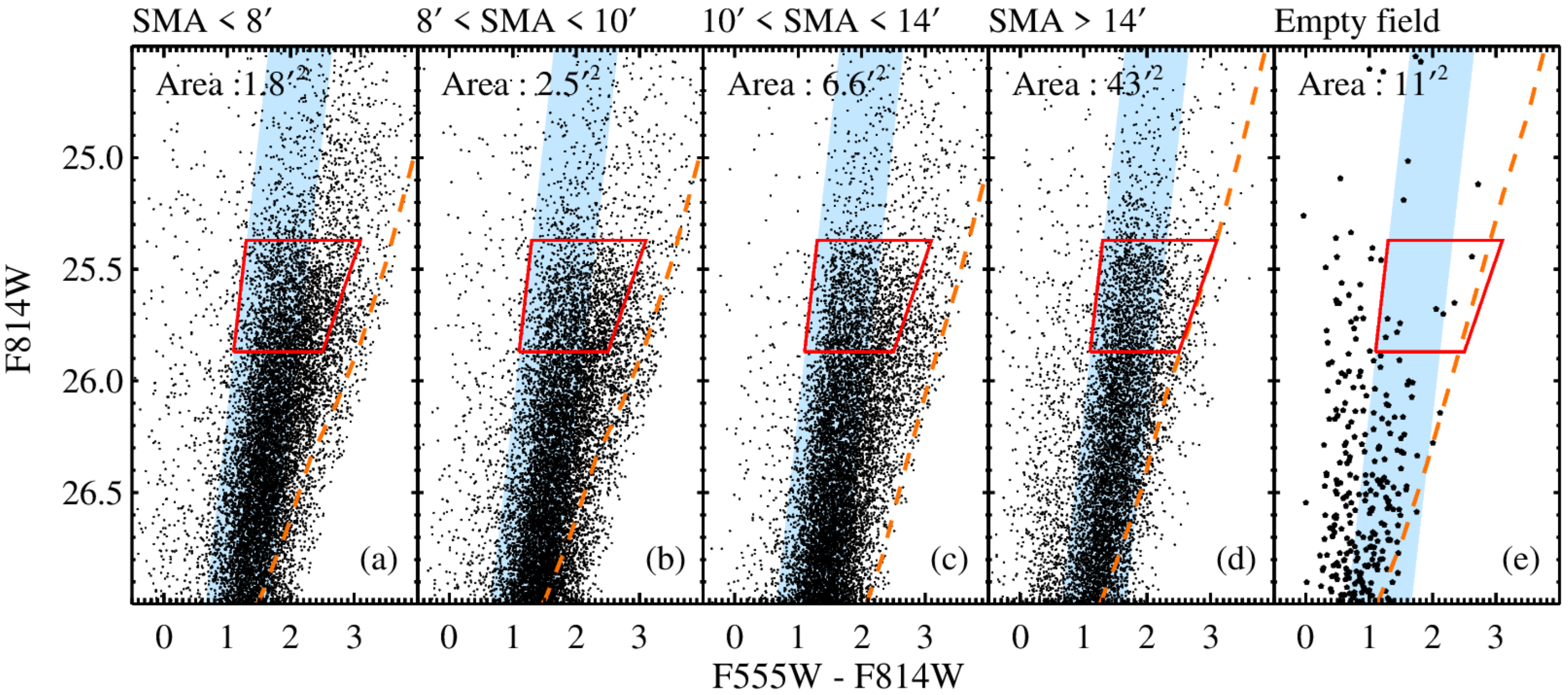}
    \caption{
    CMDs of resolved stars in the 
    mosaic field of \gal. 
    The panels step out in radial bins in SMA for panels (a) to (d).
    Panel (e) shows a CMD of a relatively blank field that we used for visualizing the extent of background source contamination, especially in and around the portion of the CMD used to detect and measure the TRGB.
    Shaded regions represent selection bins for the blue RGB stars.
    The red polygon denotes the selection region used for the star count profile.
    The sequence of bright RGB stars (inside the red boxes) narrows and shifts blueward with increasing radius from the center. The outermost bin is dominated by blue (metal-poor) RGB stars.
    Dashed (orange) lines show the 50\% completeness level for each field.}
    \label{fig:cmd_5fields}
\end{figure}
 
In \autoref{fig:cmd_5fields}, we show CMDs of the mosaic field sub-divided into four regions based on the semi-major axis (SMA) radial distance from the center of \gal: (a) $SMA\leq8\arcmin$, (b)  $8\arcmin < SMA \leq 10\arcmin$, (c) $10\arcmin < SMA \leq 14\arcmin$, and (d) $SMA > 14\arcmin$.
Each CMD has approximately the same number of stars in the shaded-blue region representing our RGB domain; there are approximately 3,000 RGB stars within one-magnitude fainter than the TRGB in each CMD.
The individual field areas are not the same; the outer regions covering wider/larger areas, as annotated at the top of the individual panels.
The CMDs of \gal fields show a gradual change, wherein the sequence of the bright RGB stars near the tip (i.e., stars inside the red box) gets narrower in moving from the inner to the outer regions.
The outermost region is dominated and well defined by the blue (metal-poor) RGB stars, similar to the RGB populations in stellar halos of other nearby disk galaxies.

\autoref{fig:cmd_5fields}(e) shows a CMD of a `blank field' used for the assessment of non-\gal background sources\footnote{The position of this field is R.A. = 12:05:45.29, and Decl. = +49:10:53.4, which is about  $2.9\deg$ away from the \gal center. This ACS field was obtained with a primary aim to study the mass structure of distant lens galaxies \citep[][PID = 10494]{hst_prop_10494},
but it is also useful as background control field for the \gal fields studied here. Exposure times are F555W in 2320s and  F814W in 2388s, similar to the mean depth of the outer regions of the mosaic field.}.
The sources in this relatively `blank field' are either foreground stars in the Milky Way or background galaxies that are sufficiently unresolved to pass through our point-source selection filtering (\autoref{sec:culls}).

We selected all sources within the red polygon shown in the panels of \autoref{fig:cmd_5fields} and plotted their radial star-count profile using filled circles in \autoref{fig:rdp_3rgbs}.
This profile can be divided into two groups: the blue RGB stars satisfying the color criteria we used for the TRGB detection (those within blue-shaded regions in CMDs and shown by blue circles in \autoref{fig:rdp_3rgbs}) and the remaining of the sources that are redder than the blue criteria (red circles).
The profiles in \autoref{fig:rdp_3rgbs} are corrected for (a) photometric incompleteness (derived from the artificial star data) and (b) the background contamination from the `blank field' observations (\autoref{fig:cmd_5fields}(e)).
The thick gray line is the  $B$-band integrated light profile of \gal from \citet{Watkins_2016}.
We took the $B$-band profile for the east side of \gal (mean of the green and red lines in Figure~6 of their paper) to get a similar spatial sampling as our RGB counts.
The $B$-band integrated light is scaled to the RGB density using Padova stellar models \citep{Bressan_2012}; we generated a well-populated model CMD that has an isochrone-age of 10 Gyr and a metallicity of $[Fe/H] = -1.0$ and assumed a Chabrier initial mass function \citep{Chabrier_2003} from which we derived a relation between the number of RGB stars obeying our selection criteria and its $B$-band total luminosity.
The integrated light profile ends 
at $SMA\sim21\arcmin$ where the surface brightness reaches $\mu_B\sim29$ mag arcsec$^{-2}$. 
Beyond this radius, the RGB profiles continues out to $SMA\sim28\arcmin$.
There is a slight systematic offset of $\sim$0.4~mag between the integrated light and the total RGB counts, which could be due to the presence of the younger stellar populations (mostly AGB stars) as shown in CMDs.

Acknowledging this expected offset, the profiles from the integrated light and total RGB starcount (filled circles) show similar overall trends: the density gradient becomes shallower with increasing radial distance.
This transition is mostly due to the blue (metal-poor) RGB population (blue circles), given that the metal-rich RGB stars (red circles) show a more rapid exponential decay with galacto-centric distance.

We fit the blue and red RGB profiles separately with Sersic laws, obtaining Sersic indices of  $n_{blue RGB} =   3.6\pm1.2$  and $n_{red RGB} = 1.1\pm0.2$, respectively. 
Thus, the blue RGB stars have a spatial distribution similar to other stellar halos, while the red RGB stars follow the disk-like profile.
Fitting the blue RGB profile with a power-law results in a power-law index of $\alpha = -3.5\pm0.1$ that is consistent with the slope of stellar halos in nearby MW mass disk galaxies ($-5.3 \leq \alpha \leq -2.7$) \citep{Harmsen_2017}.
Fitting the red RGB profile with an exponential law, we obtain an exponential disk scale length of $h_d = 2\farcm5 \pm 0\farcm1$ ($5.5\pm0.2$~kpc), which is similar to the disk scale length of about 6~kpc, as measured by \citet{Watkins_2016}.

The bottom panel of \autoref{fig:rdp_3rgbs} plots the color of  stars from the selection polygon in \autoref{fig:cmd_5fields} (grey) against SMA; the median color is computed in radial bins and plotted as filled symbols.
A negative color gradient is evident that is related to the population change such that the number of red RGB stars is declining more rapidly than the blue stars.
There is no clear boundary between the stellar disk and halo of \gal, but we infer that a region with $SMA \gtrsim16\arcmin$ is a natural point beyond which we reliably sample halo stars; both the $B$-band integrated light and the total RGB star count profiles start diverging from the extension of the inner disk profile at this radius.
For the stars interior to $SMA \gtrsim16\arcmin$ care needs to be taken in recognition of the growing contribution of disk stellar populations. Up to a point, this contribution is, at least, be minimized by judiciously using only the blue RGB stars (e.g., the shaded box in \autoref{fig:cmd_5fields}).

{\color{red}
\begin{figure}
\figurenum{A3}
    \centering
    \includegraphics[width=0.4\textwidth]{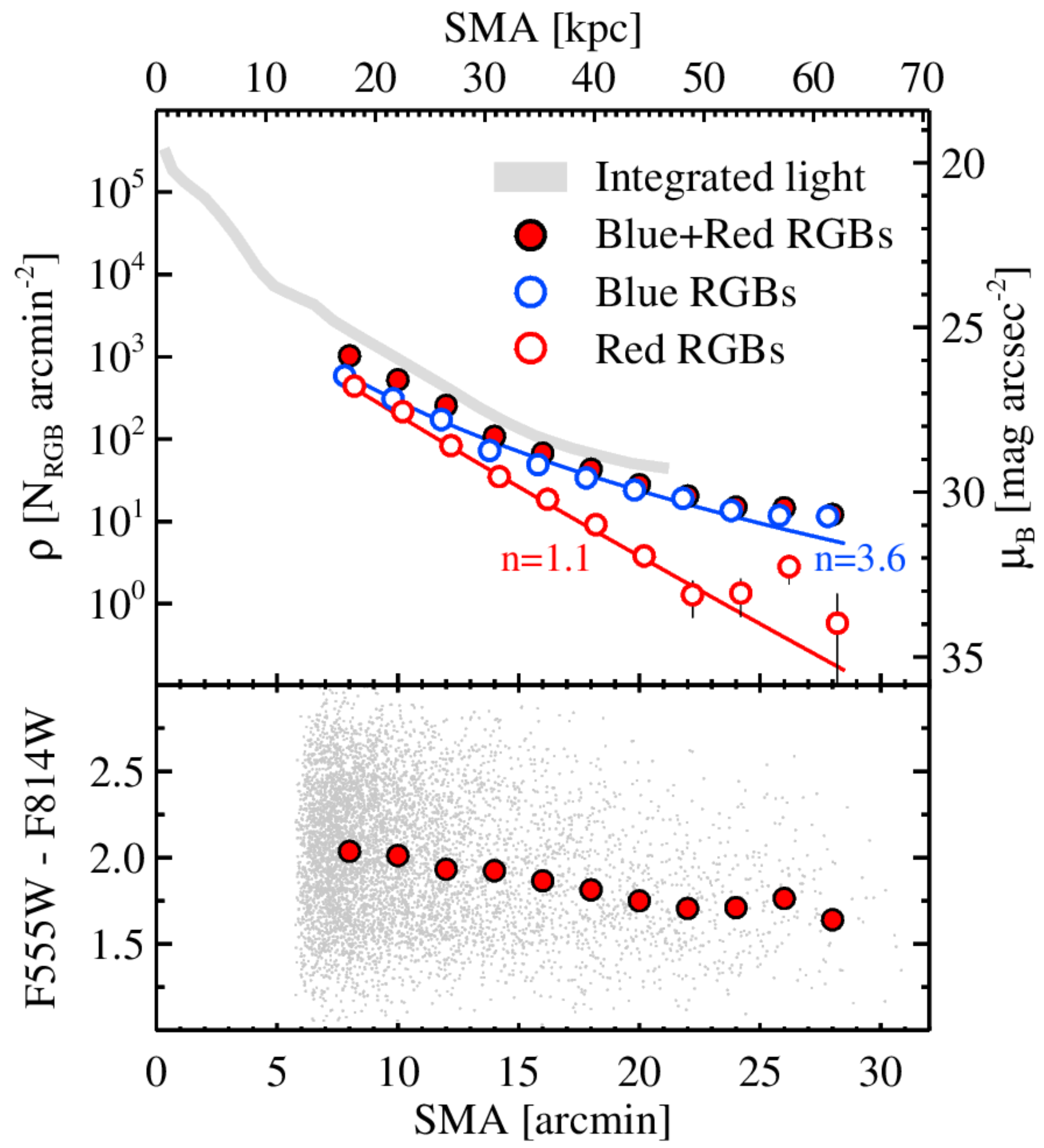}
    \caption{
    Top Panel: Radial starcount profiles of the blue RGB stars (blue circles), the red RGB stars (red circles), and their sum (filled circles), following the selection polygon in \autoref{fig:cmd_5fields}. 
    Solid lines represent fitted Sersic profiles for the blue and red RGB density profiles with the parameters specified in the text.
    The $B$-band integrated light profile from the minor axis of \gal in \citet{Watkins_2016} is shown by a thick gray line; there is a small systematic offset between this and our RGB starcounts.
    Bottom Panel: Radial color profile for the RGB stars in the selection polygon from \autoref{fig:cmd_5fields}. Individual stars and their median color are indicated by grey dots and filled (red) circles, respectively.
    }
    \label{fig:rdp_3rgbs}
\end{figure}
}

\section{An Independent Analysis of the WFPC2 Halo Field} \label{app:appc}

There are WFPC2 observations 
\citep[][PID = 9086]{hst_prop_9086} taken on the minor 
axis of \gal, where stellar disk contamination
and {\it in situ} dust extinction are each expected to be low.
The exposure times were 11,412s in F606W and 11,700s in
F814W, certainly sufficient to detect and measure the
resolved RGB population of stars at the distance of \gal.
Indeed, this field has been successfully used in two previous studies, each measuring TRGB distances. 
However, the first study opted to use aperture photometry \citep{Mouhcine_2005} that could have crowding issues.  And the second study used the $T$ magnitude system \citep{Mager_2008} for the tip measurement, which is distinct from the F814W magnitude systems used in this work.

We have independently reduced the WFPC2 data and determined the TRGB. We downloaded the science extension (\_c0m) of the WFPC2 data and processed them using the DrizzlePac to get a fine alignment solution and better data quality extensions (\_c1m).
We then used DOLPHOT to derive magnitudes from PSF fitting and leave the magnitudes in the F814W magnitude system.
The input parameters we used for the photometry are the same as those in the DOLPHOT/WFPC2 User’s guide.
The point source selection was made using the photometric diagnostic parameters: $-0.5 < {\rm Sharpness_{F814W}} \leq 0.5$, $S/N_{\rm F606W} > 3.0$, $S/N_{\rm F814W} > 3.0$, and $\rm Type=1$.

\begin{figure}
\figurenum{A4}
\centering
    \includegraphics[width=0.7\textwidth]{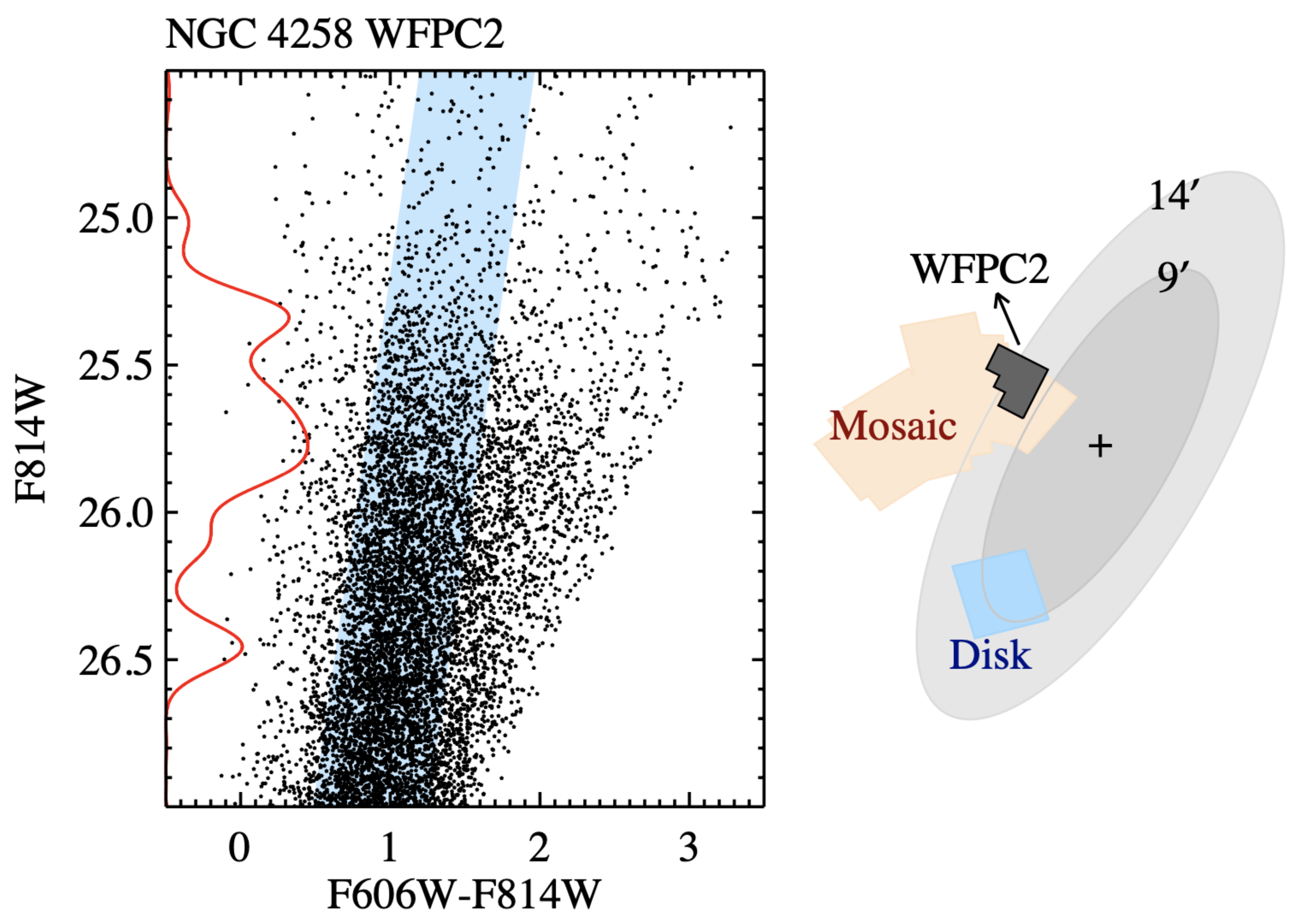}
    \caption{
    CMD for resolved stars in the WFPC2 field of \gal. The edge-detection filter response (red line) derived from the blue RGB stars (shown in the upward slanting, blue-shaded region) finds a visible peak at $F814W = 25.34 \pm 0.1$~mag, which is identified as the TRGB. Locations of the WFPC2 field (filled black footprint) with respect to the NGC 4258 ``Disk'' field (blue square) and the mosaic fields (light yellow) are sketched on the right. The filled grey ellipses show the main optical body of the maser-host galaxy \gal.}
    \label{fig:wfpc2}
\end{figure}

The CMD of the selected point sources in the three wide field chips is shown in \autoref{fig:wfpc2}.
The edge detection algorithm applied to the blue RGB stars (shaded region) finds a peak response at $F814W = 25.34\pm0.099$ mag, which we identify with the TRGB.
Here the cumulative error is a conservative estimate, derived from the following individual sources: tip detection ($\sigma =0.05$~mag), aperture correction ($\sigma =0.05$~mag), and WFPC2 zero-point ($\sigma =0.07$~mag).
This pure WFPC2 data-based measurement is now statistically consistent with our primary result based on the mosaic field, indicating that there is nothing intrinsically flawed with this field or its placement. 

We recall from the main text that the  prior measurements on this field are $I_{\rm TRGB}=25.25^{+0.13}_{-0.02}$~mag and $I_{\rm TRGB}=25.22\pm0.09$~mag from \citet{Mouhcine_2005} and $T_{\rm TRGB} = 25.20 \pm 0.06$ mag from \citet{Mager_2008}. 
The prior results are systematically brighter, but are also likely to be in magnitude systems that are distinct from the native ``flight magnitude'' system of HST adopted here.

\end{document}